\documentclass[preprint]{revtex4}
\usepackage{graphicx}
\usepackage{subfigure}
\usepackage{epsfig}
\usepackage[usenames]{color}

\begin{document}

\title{The quark gluon plasma equation of state and the expansion of the early Universe}

\author{S. M. Sanches Jr.\dag\, F. S. Navarra\dag\ and D. A. Foga\c{c}a\dag\ }
\address{\dag\ Instituto de F\'{\i}sica, Universidade de S\~{a}o Paulo\\
Rua do Mat\~ao Travessa R, 187, 05508-090, S\~{a}o Paulo, SP, Brazil}

\begin{abstract}

Our knowledge of the equation of state of the quark gluon plasma has been continuously growing due to the experimental results
from heavy ion collisions, due to recent astrophysical measurements and also due to the advances in lattice QCD calculations. The new findings
about this state may have
consequences on the time evolution of the early Universe, which  can estimated by solving the Friedmann equations.
The solutions of these equations give the time evolution of the energy density and also  of the temperature in the beginning of
the Universe.  In this work we compute the time evolution of the QGP in the early Universe, comparing several equations of state,
some of them based on the MIT bag model (and on its variants) and some of them based on  lattice QCD calculations.
Among other things, we investigate the effects of a finite baryon  chemical potential  in the evolution
of the early Universe.

\end{abstract}

\maketitle

\section{Introduction}

In the last ten years relativistic heavy-ion collision experiments have provided us with  information about
the properties of matter in the early Universe (at the time when its age was less than  $10$ microseconds and its temperature was higher than $150$ $MeV$).
It is believed that, during this period, the Universe was formed by a hot phase of deconfined quarks and gluons, i.e., a quark gluon plasma (QGP).
In parallel with these experimental developments there has been a significant progress on the  theoretical side, coming from the numerical simulation of
finite temperature QCD on a lattice. The new findings about the nature  of the QGP motivate us to investigate their consequences in the primordial
Universe. This can be done by solving the Friedmann equations, which allow us to determine the precise time evolution of the thermodynamic quantities
in the early Universe.

Previous works along this line and with the same motivation  already exist in the literature. For a review see, e.g., \cite{cosmo1} and for recent
papers on the subject see \cite{guardo,florko} and references therein. Most of these works focused on the nature of the phase transition from the QGP to the
hadron gas. There are exotic phenomena associated with the order of the phase transition. In \cite{florko}  a  realistic EOS was used in cosmological
calculations. In this EOS the transition was actually a crossover and not a first order transition as commonly believed until recent years.
The results showed a very smooth time dependence of various thermodynamic quantities and suggested indirectly that there are small
chances for the observation of various exotic phenomena such as quark nuggets, strangelets, cold dark matter clumps, etc. Such phenomena are associated
typically with  first order phase transitions. Apart from these exotic phenomena, changing the equation of state we change the space-time evolution of
the early Universe and this (specially when there is a phase transition) will change the emission of gravitational waves, as pointed out in \cite{lugo},
\cite{domi1}, \cite{domi2} and very recently in \cite{tigran}.  In \cite{schet} the authors have considered different EOS for the QGP. They computed the energy-momentum
tensor, performed a Fourier transform from the configuration to the momentum space and then, using a textbook formula from \cite{wein},
computed the wave spectrum, i.e., the energy density radiated in gravitational  waves as a function of the wave frequency.
Different EOS yield different spectra and the difference is larger for higher frequencies ( $\nu > 10^{-9}$ $Hz$). In the region $\nu > 10^{-5}$ $Hz$,
depending on details of the phase transition,  the differences can be of orders of magnitude in the spectrum. The differences might be detectable.
In \cite{lugo} the authors show that the  eLISA/NGO (New Gravitational wave Observatory)  planned for the next years (and  also the Big Bang
Observatory (BBO))  will be able to measure the gravitational radiation in the frequency region relevant to the  quark gluon plasma physics. They show
how changes in the spectrum  (due to  fluctuations in temperature and fluid velocity) could be observed by eLISA/NGO in the frequency
region  $\nu > 10^{-5}$ $Hz$. The authors have considered the recently published eLISA/NGO sensitivity curve.

In the early Universe the baryon chemical potential was small (and usually neglected in
cosmological calculations) but we do not know exactly how small.  Moreover there may have been  fluctuations in the chemical
potential  associated with the anisotropy of positively and negatively charged
 particles in the QGP phase, as pointed out in \cite{sany}.
It is therefore interesting to estimate the effects of a non-vanishing chemical potential on the solution of the Friedmann equations.

We believe that the Universe is homogeneous and isotropic \cite{wein}. This statement implies that the space-time can be parametrized by the
Friedmann-Lema\^{i}tre-Robertson-Walker (FLRW) metric which, inserted into the Einstein equations yield the Friedmann equations \cite{wein}.
From these latter we can derive the following time evolution equation \cite{guardo,florko,cosmo2}:
\begin{equation}
-\frac{d\varepsilon}{3 \sqrt{\varepsilon} \left( \varepsilon + p \right)} = \sqrt{\frac{8 \pi G}{3}} dt
\label{eq:time-evol-E/V}
\end{equation}
which allows us  to find the temporal evolution of the energy density $\varepsilon$ once we know  $p \equiv p(\varepsilon)$.
In this work   we solve  numerically the  equation above using some recently proposed equations of state  and
compute the time evolution of some thermodynamical quantities in the early
Universe, such as the  energy density, pressure, temperature and sound speed.  We also study the time evolution of these quantities
at finite chemical potential.  In what follows we will use natural units $\hbar=c=k_{B}=1$.

\section{The QGP equation of state}

One of the first equations of state of the quark gluon plasma was the one derived from the MIT bag model \cite{mit}. Because of its simplicity it
has been widely used
in astrophysics and cosmology. Even today it remains a baseline. As discussed in the introduction, we want to study how  changes in the EOS affect the time
evolution of the primordial QGP. In doing so, we will take the MIT EOS as a reference,  with which results obtained with the other equations of state will be
compared. In view of the most recent lattice results \cite{ratti,fodor14} one might argue that we already know the true equation of state of the QGP and there is
no need to test any other candidate. In fact, lattice calculations are a work in progress and there is always some technical aspect that might be improved. Moreover
we can not rely on the results of one single group and confirmation by other groups  is required. Especially in the case of finite chemical potential,
results obtained
with different prescriptions should be compared. Even after all these improvements, when the "final equation of state" would  become available, studies as the one
performed in the present work would still be important for, at least, two reasons. First, because we want to determine the role played separately by each ingredient
(e.g. the baryon chemical potential, non-perturbative component of the pressure,...etc) in the expansion. Second, because the QGP simulated on the
lattice may be a bit
different from the plasma in the early Universe. The latter may contain a significant leptonic component, which might affect its thermodynamical properties.
In view of
the remarks made above, we keep an open mind on the subject and consider that the QGP EOS is still preliminary. In this section we introduce and discuss
six different EOS with different properties, each of them having some phenomenological support, either from relativistic heavy ion physics or from the
study of compact stars. We believe that this set of equations of state gives a good idea of the sistematic uncertainties involved but we are aware that there many
other EOS which might be considered in our study \cite{others}.

\subsection{The MIT bag model}
\vspace{0.5cm}

In the MIT picture quarks and gluons move freely inside the "bag" and the deconfined phase can be formed by compressing the bags against each other. Then, at these
high baryon number and energy densities, the plasma constituents are free to move through large spatial regions. A higher temperature enhances quark-antiquark pair
creation. Even in
these conditions there is still some non-perturbative component represented by the bag constant.
The energy density and pressure are given respectively by \cite{mit,we10}:
\begin{equation}
\varepsilon={\frac{37\pi^{2}}{30}} \,T^{4}
+\mathcal{B}
\hspace{1.5cm} \textrm{and} \hspace{1.5cm}
p={\frac{37\pi^{2}}{90}} \,T^{4}
-\mathcal{B}
\label{eandpmit}
\end{equation}
where $\mathcal{B}$ is  the bag constant.  Considering these thermodynamical quantities as functions of time, we have from (\ref{eandpmit}) the following
relation:
\begin{equation}
p [\varepsilon(t)] = \frac{1}{3} \left[ \varepsilon(t) - 4 \mathcal{B} \right]
\label{eq:EoS-MIT-p(e)}
\end{equation}
A particular case, considering only free gluons is described by \cite{mit,we10}:
\begin{equation}
{\varepsilon}_{g}={\frac{8\pi^{2}}{15}} \,T^{4}
+\mathcal{B}
\hspace{1.5cm} \textrm{and} \hspace{1.5cm}
{p}_{g}={\frac{8\pi^{2}}{45}} \,T^{4}
-\mathcal{B}
\label{eandpmitg}
\end{equation}
which also satisfies (\ref{eq:EoS-MIT-p(e)}). The above expressions imply that, apart from a background constant, quarks and gluons are
free. However the RHIC measurements of  elliptic flow \cite{flow} gave us convincing evidence that a QGP at temperatures around the critical
temperature $T_c$ (and up to two or three times $T_c$) is still a strongly interacting system. Moreover, the measurements of Shapiro delay
in the binary millisecond pulsar PSR J1614-2230 \cite{demo} imply the existence of  compact objects with two solar masses. If we interpret
these objects as quark stars, made of cold QGP, in order to reach the required masses we need to go beyond the MIT free gas picture and introduce
a strong interaction between the quarks. There are already several attempts to modify the bag model \cite{alford,giacosa}. In the next subsections
we will consider the models which try to adapt the MIT bag model, making it compatible with the existing lattice QCD data.

In Ref. \cite{begun} the proposed modifications of the bag model were:
a reduction in the Stephan-Boltzmann constant; the introduction of  another temperature dependent term (linear or
quadratic)  in the pressure and also in the energy density; a bag constant term with negative sign.
These features were found to be necessary to describe the lattice QCD data \cite{rede1} and they resulted in two viable simple
models, which are listed below.

\subsection{ Model 1}
\vspace{0.5cm}

The pressure and energy density as function of temperature are given respectively by:
\begin{equation}
p_{1} = \frac{\sigma_{1}}{3} T^4 - A T - \mathcal{B}_{1}
\hspace{1.5cm} \textrm{and} \hspace{1.5cm}
\varepsilon_{1} = \sigma_{1} T^4 + \mathcal{B}_{1}
\label{peabag}
\end{equation}
which gives the following relation:
\begin{equation}
p_{1} [\varepsilon_{1}(t)] = \frac{1}{3} \left[ \varepsilon_{1}(t) - 4 \mathcal{B}_{1} \right] - A \left[ \frac{\varepsilon_{1}(t) -
\mathcal{B}_{1}}{\sigma_{1}} \right]^{1/4}
\label{eq:EoS-Abag-p(e)}
\end{equation}
with the parameters from \cite{begun}: $\sigma_{1} = 4.73 $, $ A = 3.94 \,{T_c}^3 $ and $ \mathcal{B}_{1}
= - 2.37 \, {T_c}^4 $, where $T_c$ is the critical temperature for the QGP.

\subsection{Model 2}
\vspace{0.5cm}

The introduction of terms proportional to $T^2$ in the pressure and energy density was suggested by Pisarski \cite{pisarski}
to take into account non-perturbative effects in the QGP.
The pressure and energy density  are given respectively by:
\begin{equation}
p_{2} = \frac{\sigma_{2}}{3} T^4 - C T^2 - \mathcal{B}_{2}
\hspace{1.5cm} \textrm{and} \hspace{1.5cm}
\varepsilon_{2} = \sigma_{2} T^4 - C T^2 + \mathcal{B}_{2}
\label{pecbag}
\end{equation}
and so:
\begin{equation}
p_{2} [\varepsilon_{2}(t)] = \frac{1}{3 \sigma_{2}} \left\{ \left[ \varepsilon_{2}(t) - 4 \mathcal{B}_{2} \right] - C \left[ C + \sqrt{C^2
+ 4 \sigma_{2} \left[ \varepsilon_{2}(t) - \mathcal{B}_{2} \right]} \right] \right\}
\label{eq:EoS-Cbag-p(e)}
\end{equation}
and again from \cite{begun} : $ \sigma_{2} = 13.01 $, $ C = 6.06 \, {T_c}^2 $ and $ \mathcal{B}_{2} = - 2.34 \, {T_c}^4 $.

\subsection{Model 3}
\vspace{0.5cm}

In spite of the success in reproducing the lattice results, the models described  in the previous subsections  do not have a clear connection with the
QCD dynamics.  The model  presented below was proposed in  \cite{we11} and it  was applied to study the cold and dense quark gluon plasma in the inner
core of neutron stars \cite{we12}. In \cite{we11} we start from the QCD Lagrangian and split the gluon field into low and high momentum modes. The former
are rewritten in terms of the gluon condensates (which are assumed to have a residual non-vanishing value in the QGP phase) and the latter are replaced by
classical fields, in the same way as it is done in relativistic mean field models of nuclear matter.

Using the effective Lagrangian derived in   \cite{we11} and repeating the steps of the finite temperature formalism developed in \cite{furn} we obtain the
following expression for the pressure:
\begin{equation}
p_{3}={\frac{3{g}^{2}}{16{m_{g}}^{2}}}{\rho}^{2}-\mathcal{B}_{QCD}
+\sum_{f}\,{\frac{\gamma_{f}}{6\pi^{2}}}\int_{0}^{\infty} dk \,{\frac{{k}^{4}}{\sqrt{m_{f}^{2}+k^{2}}}}\, \Big(d_{f}+\bar{d}_{f} \Big)
+{\frac{\gamma_g}{6\pi^{2}}}\int_{0}^{\infty} d{k}\,k^{2}\,(e^{k/T}-1)^{-1}
\label{pressureaalmendfbzfinalmesmo}
\end{equation}
and the energy density:
\begin{equation}
\varepsilon_{3}={\frac{3{g}^{2}}{16{m_{g}}^{2}}}{\rho}^{2}+\mathcal{B}_{QCD}
+\sum_{f}\,{\frac{\gamma_{f}}{2\pi^{2}}} \int_{0}^{\infty} dk \, k^{2}\, {\sqrt{m_{f}^{2}+k^{2}}}\,\Big(d_{f}+
\bar{d}_{f}\Big)
+{\frac{\gamma_g}{2\pi^{2}}}\int_{0}^{\infty} d{k}\,k^{2}\,(e^{k/T}-1)^{-1}
\label{endtotalalmosthafimbzendparcint}
\end{equation}
where the Fermi distribution functions are given by:
\begin{equation}
d_{f}\equiv{\frac{1}{1+e^{(\mathcal{E}_{f}-\nu_{f})/T}}}
\hspace{2.5cm} \textrm{and} \hspace{2.5cm}
\bar{d}_{f}\equiv{\frac{1}{1+e^{(\mathcal{E}_{f}+\nu_{f})/T}}}
\label{dists}
\end{equation}
The energy of the quark of flavor $f$ is given by $\mathcal{E}_{f}=\sqrt{m_{f}^{2}+k^{2}}$
and $\nu_{f}$ is the corresponding  chemical potential.  The quark density is given by:
\begin{equation}
\rho=\sum_{f}\,{\frac{\gamma_{f}}{(2\pi)^{3}}}\int d^{3}k \Big(d_{f}-\bar{d}_{f} \Big)
\label{rhoquarksfromomegaendbz}
\end{equation}
For the sake of simplicity, we  consider only two quark flavors, up and down, with equal masses, and the regime of high temperature given
by $T>>\nu_{f}$, $T>>m_{f}$ and $\mathcal{E}_{f}/T>\nu_{f}/T$. The statistical factors  are given by
$\gamma_g=2\textrm{(polarizations)}\times 8\textrm{(colors)}=16$ for gluons and
$\gamma_f=2\textrm{(spins)}\times 3
\textrm{(colors)}=6$ for each quark species.
The condition of high temperature $T>>m_{f}$ allows us to solve analytically the integrals in (\ref{pressureaalmendfbzfinalmesmo}),
(\ref{endtotalalmosthafimbzendparcint}) and (\ref{rhoquarksfromomegaendbz}) using the formulas given in \cite{tooper}.  Then
(\ref{pressureaalmendfbzfinalmesmo}) and (\ref{endtotalalmosthafimbzendparcint}) become:
\begin{equation}
p_{3}={\frac{3{g}^{2}}{16{m_{g}}^{2}}}{T^{4}}\mu^{2}
+{\frac{37\pi^{2}}{90}} \,T^{4}
-\mathcal{B}_{QCD}
+{\frac{1}{2}} \,T^{2}\,{\mu}^{2}
\label{phigh}
\end{equation}
and
\begin{equation}
\varepsilon_{3}={\frac{3{g}^{2}}{16{m_{g}}^{2}}}{T^{4}}\mu^{2}
+{\frac{37\pi^{2}}{30}} \,T^{4}
+\mathcal{B}_{QCD}
+{\frac{3}{2}} \,T^{2}\,{\mu}^{2}
\label{ehigh}
\end{equation}
where $\mu \equiv \nu_{u}=\nu_{d}$ is the chemical potential. Expressions (\ref{phigh}) and (\ref{ehigh}) are valid at very
high temperature and finite chemical potential. Setting $\mu=0$ we recover the equation of state of the  bag model \cite{mit,we10}.
The new contribution  are the term proportional to $\mu^{2}$ and the  term proportional to $g/{m_{g}}^{-2}$ \cite{we11}. The model has three
parameters, $g$, $m_g$ and $\mathcal{B}_{QCD}$ which  will be discussed later.

\subsection{ Model 4}
\vspace{0.5cm}

The equation of state below was derived in \cite{castor}, where deconfined matter in $SU(3)$ pure gauge theory was treated  as an ideal gas with
quasi-particle modes, which have a temperature-dependent mass given by $m(T)$.
This EOS reproduces the Lattice results, such as the trace anomaly $(\varepsilon-3\, p)/T^{4}$,
as can be seen in \cite{castor}.  The pressure and energy density as functions of the temperature are given by:
\begin{equation}
p_{4}={8\, T^{2} \, {m^2(T)} \over \pi^2}\,K_{2}\bigg[{\frac{m(T)}{T}}\bigg]
\label{lattpress}
\end{equation}
and
\begin{equation}
\varepsilon_{4}={8\, T^{2} \, {m^2(T)} \over \pi^2}\,
\left\{3 K_{2}\bigg[{\frac{m(T)}{T}}\bigg] + \left[{m(T) \over T} - \left({dm \over dT}\right) \right]
K_{1}\bigg[{\frac{m(T)}{T}}\bigg] \right\}
\label{lattener}
\end{equation}
where $K_{1,2}[x]$ are the modified Bessel functions of the second kind.
The quasi-particle mass is given by:
\begin{equation}
m(T) = {\frac{a}{({\frac{T}{T_{0}}}-1)^c}} + b\,{\frac{T}{T_{0}}}
\label{quasimass}
\end{equation}
with constants $a=0.47 \, GeV, b=0.125 \, GeV$ and $c=0.385$ as previously calculated in \cite{castor} to reproduce the trace anomaly.  The critical temperature in
\cite{castor} is $T_{0} \simeq 280 \, MeV$.

\subsection{ Model 5}
\vskip0.5cm

Here we present the equation of state obtained  in \cite{ratti,fodor14} from a lattice simulation of  $SU(3)$ QCD  at finite temperature and chemical potential
with three quark flavors ($u$, $d$ and $s$) with equal masses and gluons. At finite chemical potential the trace anomaly reads:
$$
{\frac{{\varepsilon}_{5}(T,\mu)-3\, {p}_{5}(T,\mu)}{T^{4}}}=T{\frac{\partial}{\partial T}}
\Bigg[{\frac{{p}_{5}(T,\mu)}{T^{4}}}\Bigg]+{\frac{\mu^{2}}{T^{2}}}\,\chi_{2}
$$
\begin{equation}
={\frac{{\varepsilon}_{5}(T,0)-3\, {p}_{5}(T,0)}{T^{4}}}
+{\frac{\mu^{2}}{2T}}\,{\frac{\partial \chi_{2}}{\partial T}}
\label{fundratti}
\end{equation}
At zero chemical potential the authors in \cite{ratti,fodor14} provide the following parametrization for the trace anomaly:
\begin{equation}
{\frac{{\varepsilon}_{5}(T,0)-3\, {p}_{5}(T,0)}{T^{4}}}=e^{-h_{1}/{\tau}-h_{2}/{\tau}^{2}} \cdot \Bigg[h_{0}+
{\frac{f_{0} \cdot \Big[tanh(f_{1}\cdot {\tau}+f_{2})+1\Big]}{1+g_{1} \cdot {\tau}+g_{2} \cdot {\tau}^{2}}}   \Bigg]
\label{rattipar}
\end{equation}
and also \cite{ratti}:
\begin{equation}
\chi_{2}= e^{-h_{3}/{\tau}-h_{4}/{\tau}^{2}} \cdot f_{3} \cdot \Big[tanh(f_{4}\cdot {\tau}+f_{5})+1\Big]
\label{rattiprea}
\end{equation}
with $\tau=T/200 $ MeV, where $200 $ MeV is the critical temperature.
The actual values for the dimensionless parameters are \cite{fodor14}: $h_{0} = 0.1396$,
$h_{1} = -0.1800$, $h_{2} = 0.0350$, $f_{0} = 1.05$,
$f_{1} = 6.39$, $f_{2} = -4.72$, $g_{1} = -0.92$ and
$g_{2} = 0.57$ . From \cite{ratti} we have $h_{3} = -0.5022$, $h_{4} = 0.5950$, $f_{3} = 0.1359$,
$f_{4} = 6.3290$ and $f_{5} = -4.8303$ .
The pressure is calculated from (\ref{fundratti}):
\begin{equation}
{p}_{5}(T,\mu)=T^{4}\, \int^{T}_{0} \, dT^{'} \,
{\frac{e^{-h_{1}/{\tau'}-h_{2}/{{\tau'}^{2}}}}{T^{'}}} \cdot \Bigg[h_{0}+
{\frac{f_{0} \cdot \Big[tanh(f_{1}\cdot {\tau'}+f_{2})+1\Big]}{1+g_{1} \cdot {\tau'}+g_{2} \cdot {{\tau'}^{2}}}}   \Bigg]
+{\frac{\chi_{2}}{2}}\, \mu^{2}  T^{2}
\label{rattipre}
\end{equation}
Inserting (\ref{rattipre}) into (\ref{rattipar}) we find the following expression for the energy density:
$$
{\varepsilon}_{5}(T,\mu)=T^{4} \,e^{-h_{1}/{\tau}-h_{2}/{\tau}^{\,2}} \cdot \Bigg[h_{0}+
{\frac{f_{0} \cdot \Big[tanh(f_{1}\cdot {\tau}+f_{2})+1\Big]}{1+g_{1} \cdot {\tau}+g_{2} \cdot {\tau}^{2}}} \Bigg]
+{\frac{\mu^{2}}{2}}\, T^{3} \, {\frac{\partial \chi_{2}}{\partial T}}
$$
\begin{equation}
+3\,T^{4}\, \int^{T}_{0} \, dT^{'} \,
{\frac{e^{-h_{1}/{\tau'}-h_{2}/{{\tau'}^{2}}}}{T^{'}}} \cdot \Bigg[h_{0}+
{\frac{f_{0} \cdot \Big[tanh(f_{1}\cdot {\tau'}+f_{2})+1\Big]}{1+g_{1} \cdot {\tau'}+g_{2} \cdot {{\tau'}^{2}}}}\Bigg]
+{\frac{3 \,\chi_{2}}{2 }} \, \mu^{2} T^{2}
\label{rattiener}
\end{equation}

\section{Numerical results and discussion}

In this  section we present the numerical solution of (\ref{eq:time-evol-E/V}) for the different equations of state discussed above.
We use the following initial condition:
\begin{equation}
 \varepsilon_{i} (t_i) = 10^{7} \, \mbox{MeV}/\mbox{fm}^3         \hspace{1.5cm} \textrm{at} \hspace{1.5cm}          t_{i} = 10^{-9}\ s
\label{init}
\end{equation}
Following the estimates made in \cite{tigran} we run the temporal evolution from the time of the electroweak phase transition, $t_{i} = 10^{-9}\ s $,
to  the time of the QCD phase transition, $t_{f} = 10^{-4}\ s$.
There is some uncertainty on these initial conditions but
this is not important for the present study, since we are primarily interested in determining how changes in the equation of state  affect the time
evolution of the
primordial QGP. Also, since the equations of state are different, fixing the initial energy density implies that the evolution starts at
different initial temperatures for different models. At this point one might ask: why fixing the initial energy density and not the initial temperature?
Because we want to be able to establish some connection  with the cold QGP at similar energy densities. As it was mentioned in the introduction,
from astrophysical
measurements, we can derive some constraints to the cold QGP properties.
In working with (\ref{eq:time-evol-E/V}) we use the relation:
$$
\frac{d T}{dt} = \frac{1}{(\frac{d \varepsilon}{dT})} \times \frac{d\varepsilon}{dt}
$$
and solve it for $T(t)$.

\subsection{The MIT bag model and its variants}

In  Fig. 1  we show the time evolution of  the energy density  (1a) and temperature (1b) obtained by  solving   (\ref{eq:time-evol-E/V})
with the MIT bag model equation of state of a quark - gluon gas (\ref{eandpmit}) and of a gluon gas (\ref{eandpmitg}), with the initial condition
(\ref{init}).
\begin{figure}[ht!]
\begin{center}
\subfigure[ ]{\label{fig:first}
\includegraphics[width=0.488\textwidth]{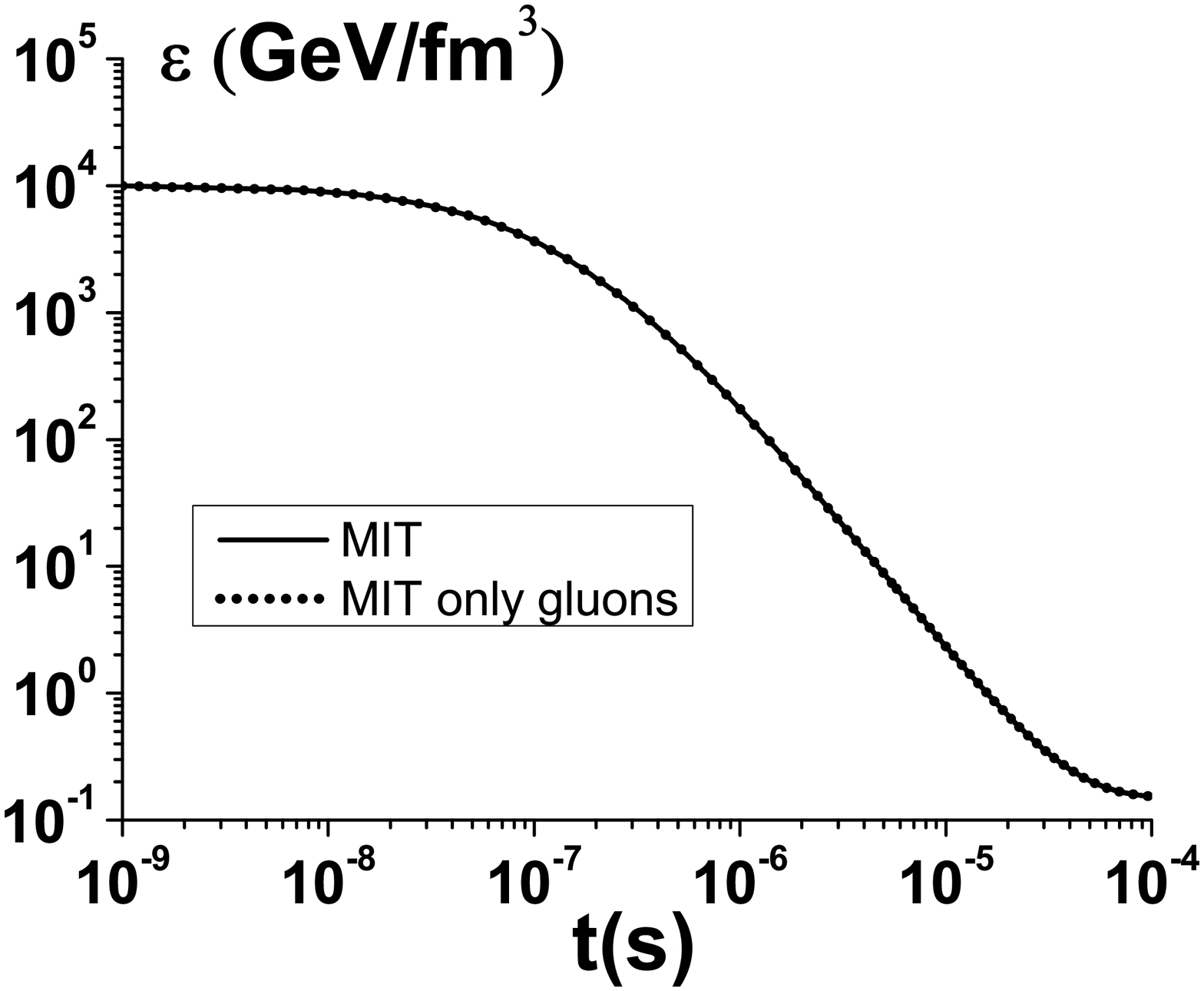}}
\subfigure[ ]{\label{fig:second}
\includegraphics[width=0.488\textwidth]{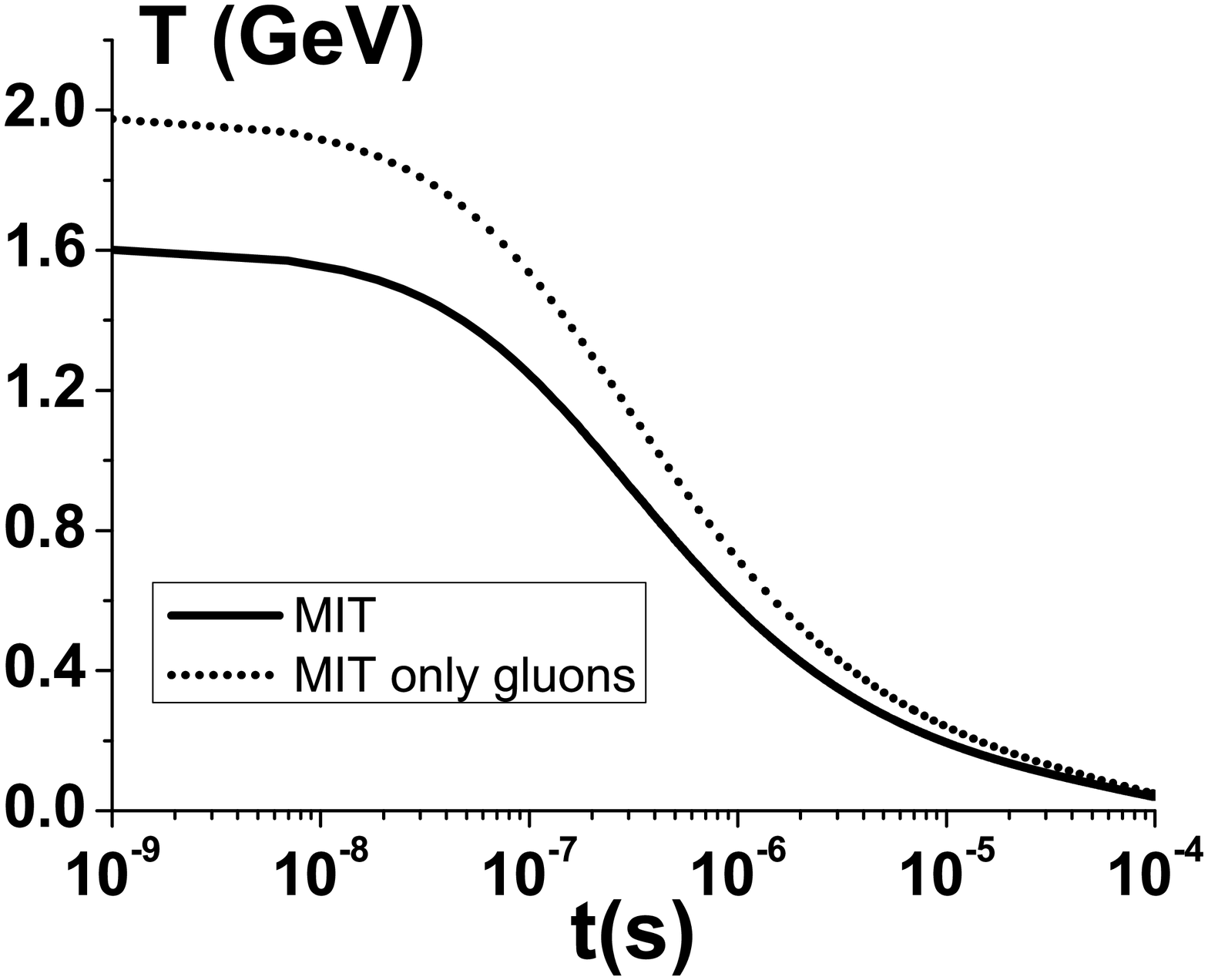}}
\end{center}
\caption{MIT bag model. (a) Time evolution of the energy density. (b) Time evolution of the  temperature.}
\label{fig1}
\end{figure}

Models 1 and 2, valid only for zero baryon chemical potential, can be regarded as the new variants of the MIT bag model which are compatible with
recent lattice QCD data. Their main feature is that they have a smaller pression and also a slower growth of the pressure with the energy density.
This leads to a smaller speed of sound, which is given by:
\begin{equation}
{c_s}^2 = \frac{\partial p}{\partial \varepsilon}
\label{eq:termo-rela-cs2}
\end{equation}
In Fig. 2 we show these quantities calculated with models 1 and 2.
\begin{figure}[ht!]
\begin{center}
\subfigure[ ]{\label{fig:first}
\includegraphics[width=0.488\textwidth]{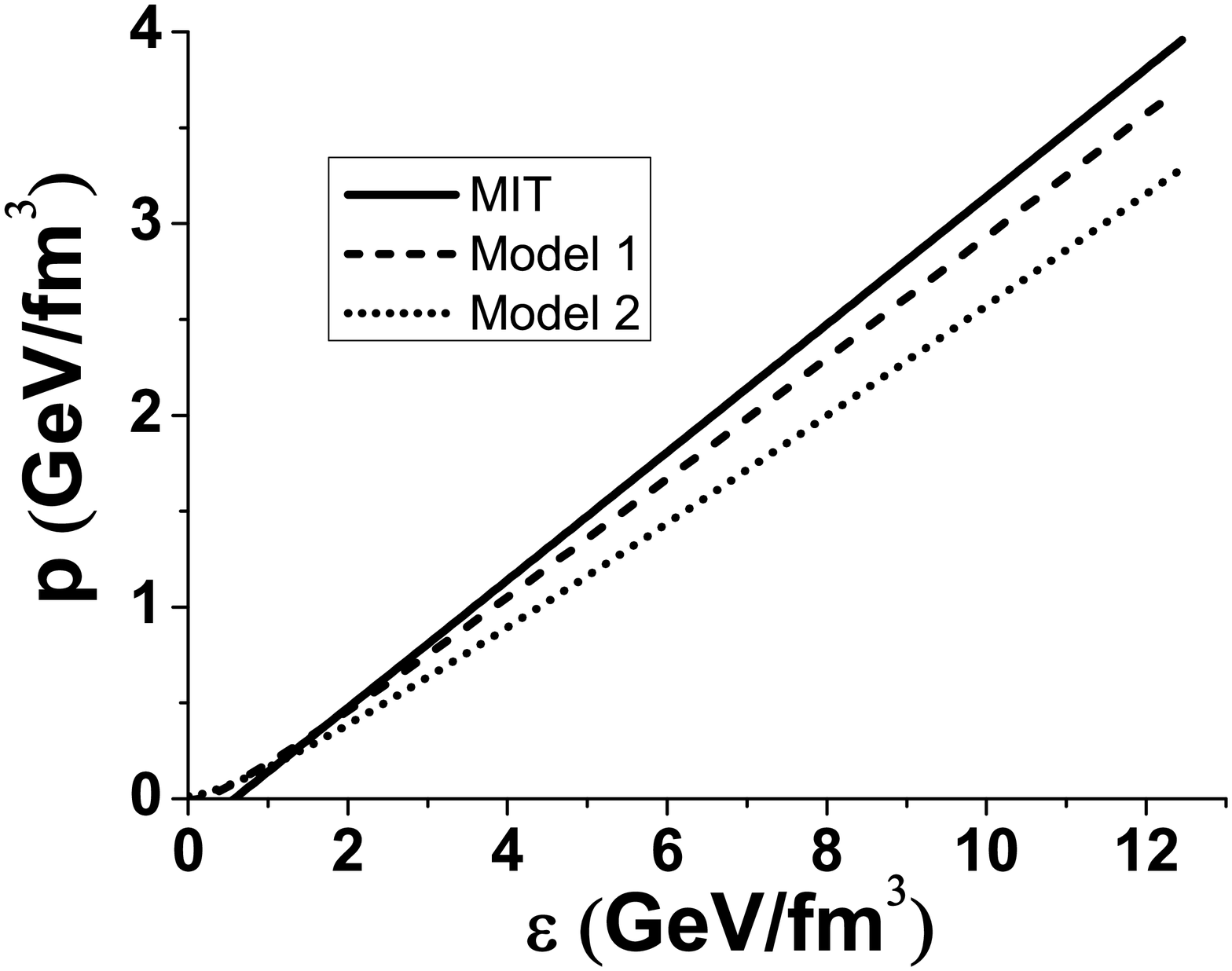}}
\subfigure[ ]{\label{fig:second}
\includegraphics[width=0.488\textwidth]{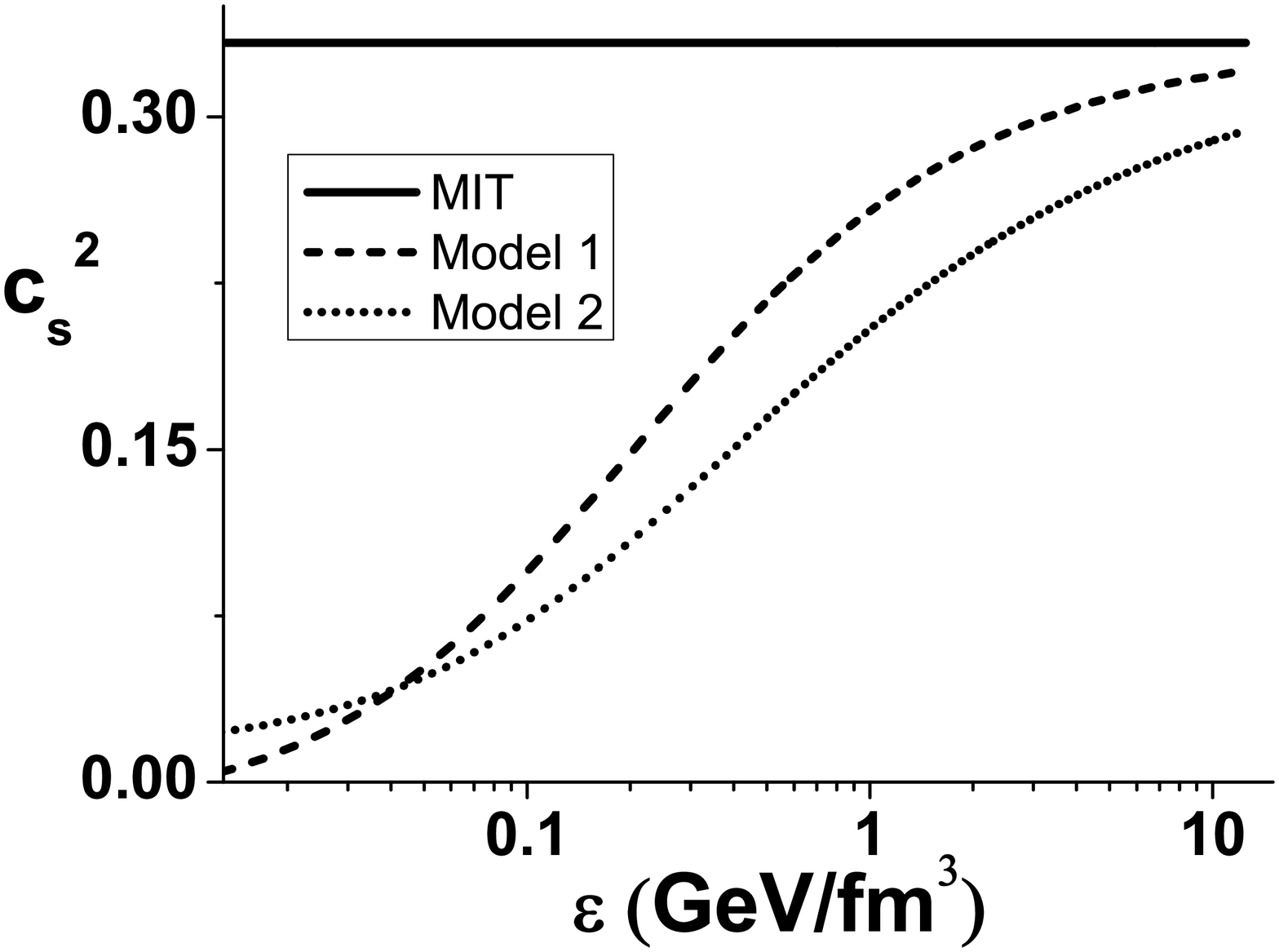}}
\subfigure[ ]{\label{fig:first}
\includegraphics[width=0.488\textwidth]{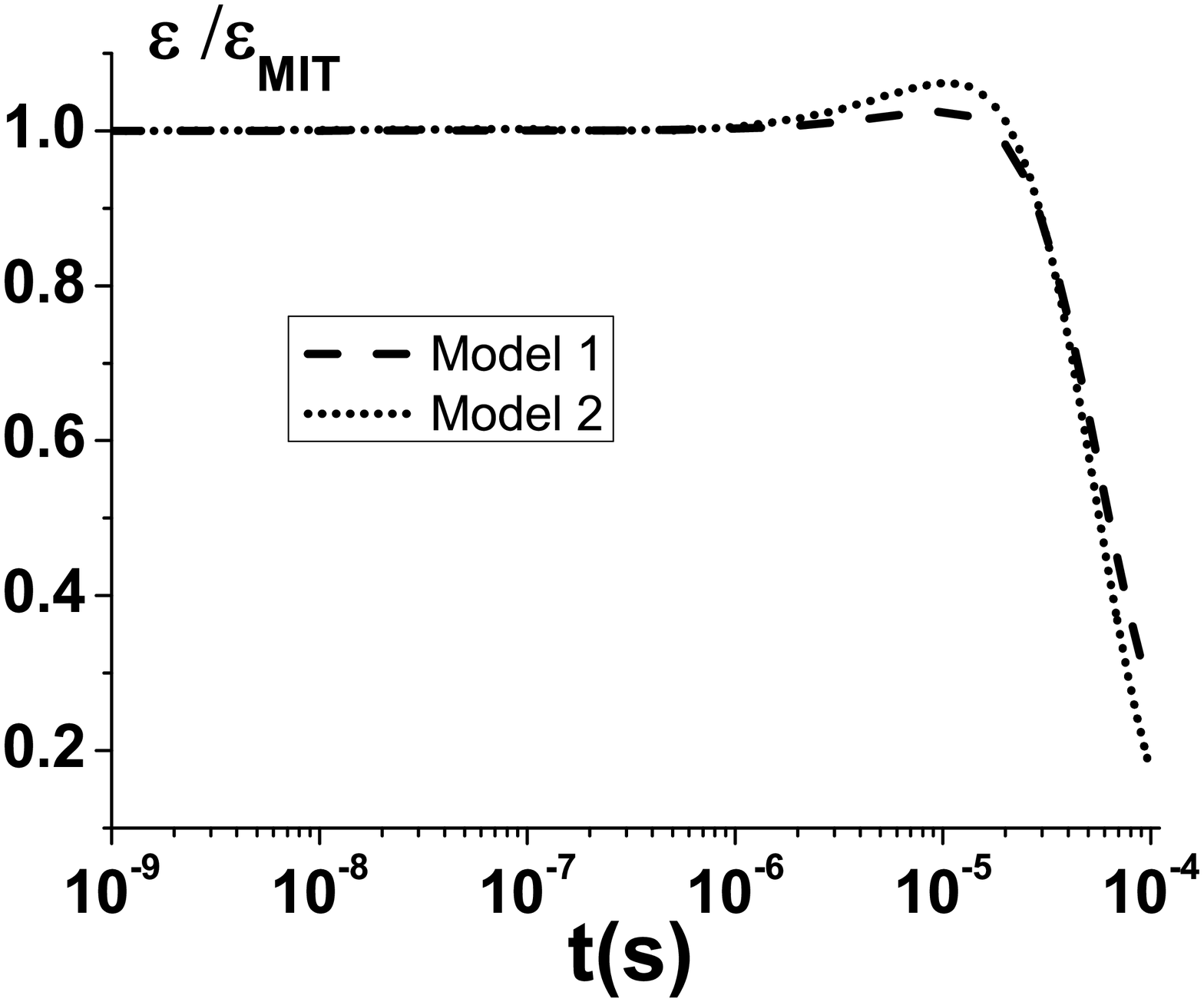}}
\subfigure[ ]{\label{fig:second}
\includegraphics[width=0.488\textwidth]{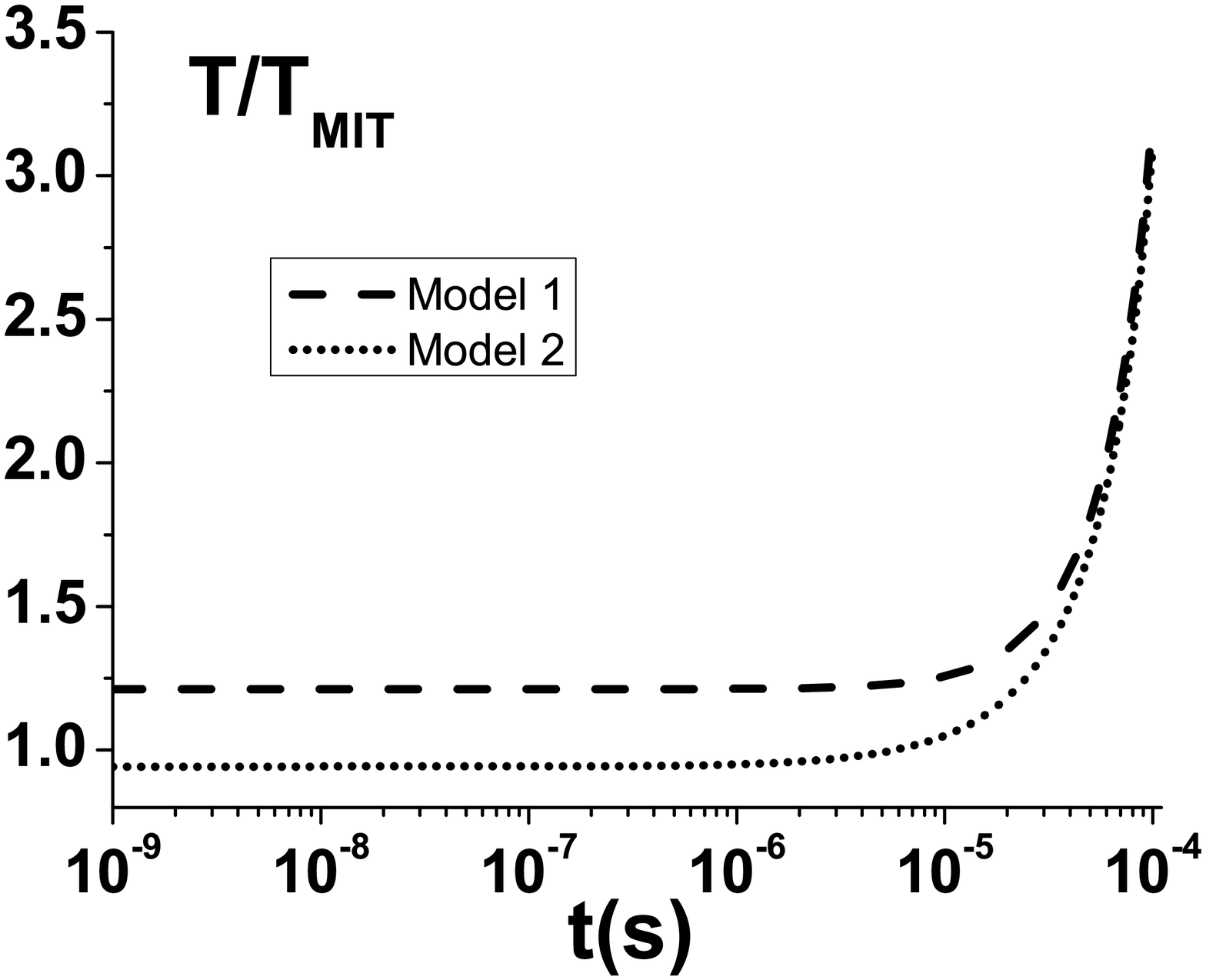}}
\end{center}
\caption{Comparison between model 1, model 2 and the MIT bag model. (a) Equation of state. (b) Speed of sound. The solid lines show the
original bag model results, calculated with $\mathcal{B}=150 \, MeV/fm^{3}$. (c) Time evolution of the energy density ratio. (d) Time evolution
of the temperature ratio.}
\label{fig2}
\end{figure}
In Fig. 2c and 2d we show the time evolution of the energy density and temperature respectively. The energies and
temperatures computed with models 1 and 2 are scaled by the corresponding quantities computed with the bag model.
Looking at these ratios, we observe that both model 1 and  model 2 have a similar behavior and that, most of the time they coincide with the bag model
predictions. However at the end of the chosen time interval, models 1 and 2 predict a much stronger energy dilution  and a much weaker cooling
than the  bag model. In fact, from Fig. 1 we see that in the interval from  $10^{-5}$ s to  $10^{-4}$ s  we expect to be sensitive to the phase transition.

\subsection{The effects of finite chemical potential}

We now repeat the  study performed above using the equation of state of model 3, Eqs. (\ref{phigh}) and (\ref{ehigh}), which allows us to consider systems with
finite chemical potential.  Model 3 has three parameters, the dynamical gluon mass, $m_{g}$,  the QCD coupling between quarks and hard gluons, $g$, and
the bag-like constant $\mathcal{B}_{QCD}$.
In Fig. 3a we show the pressure as a function of the  energy density and then we calculate the speed of sound, shown in Fig. 3b.
\begin{figure}[ht!]
\begin{center}
\subfigure[ ]{\label{fig:first}
\includegraphics[width=0.488\textwidth]{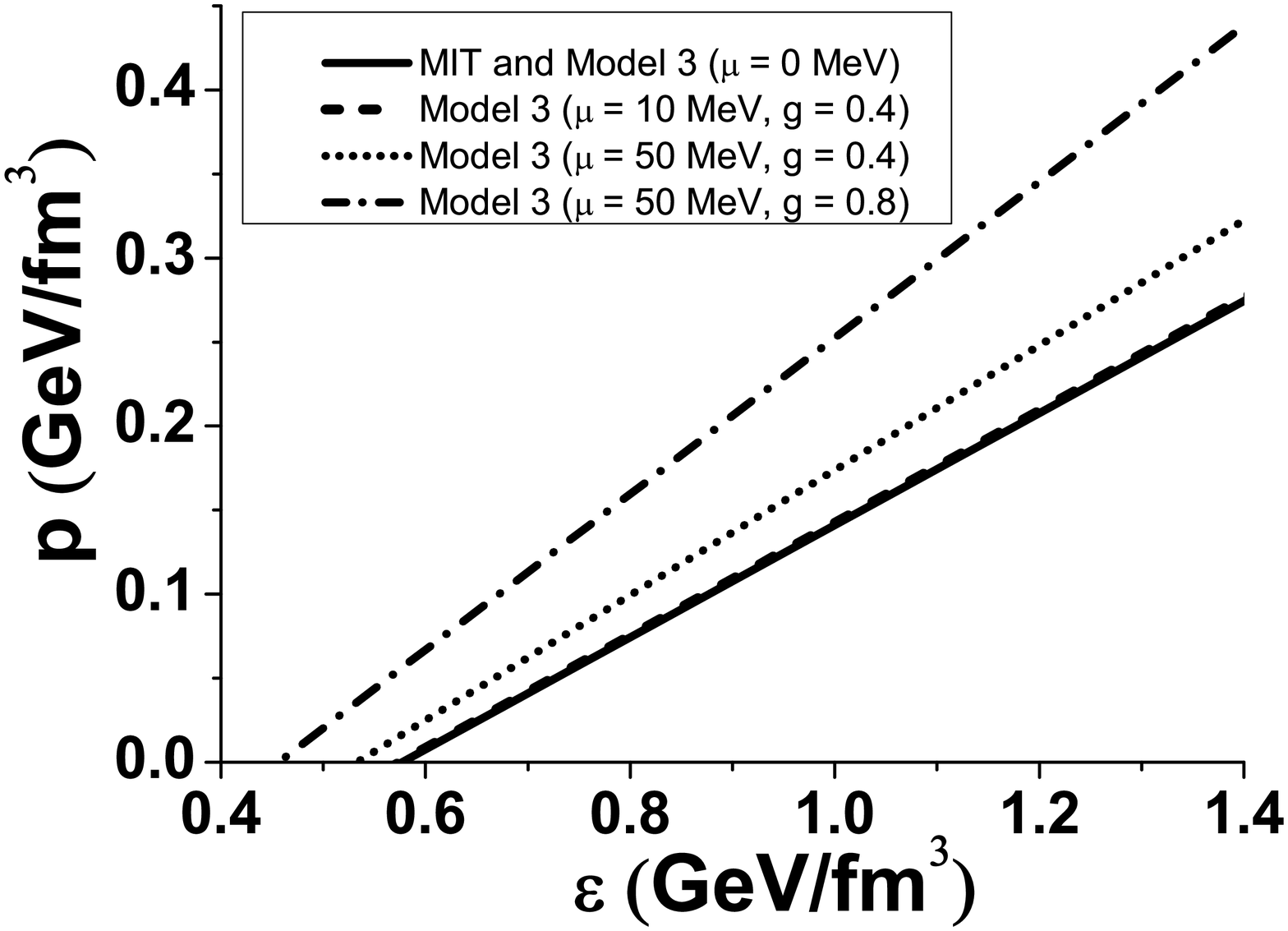}}
\subfigure[ ]{\label{fig:second}
\includegraphics[width=0.488\textwidth]{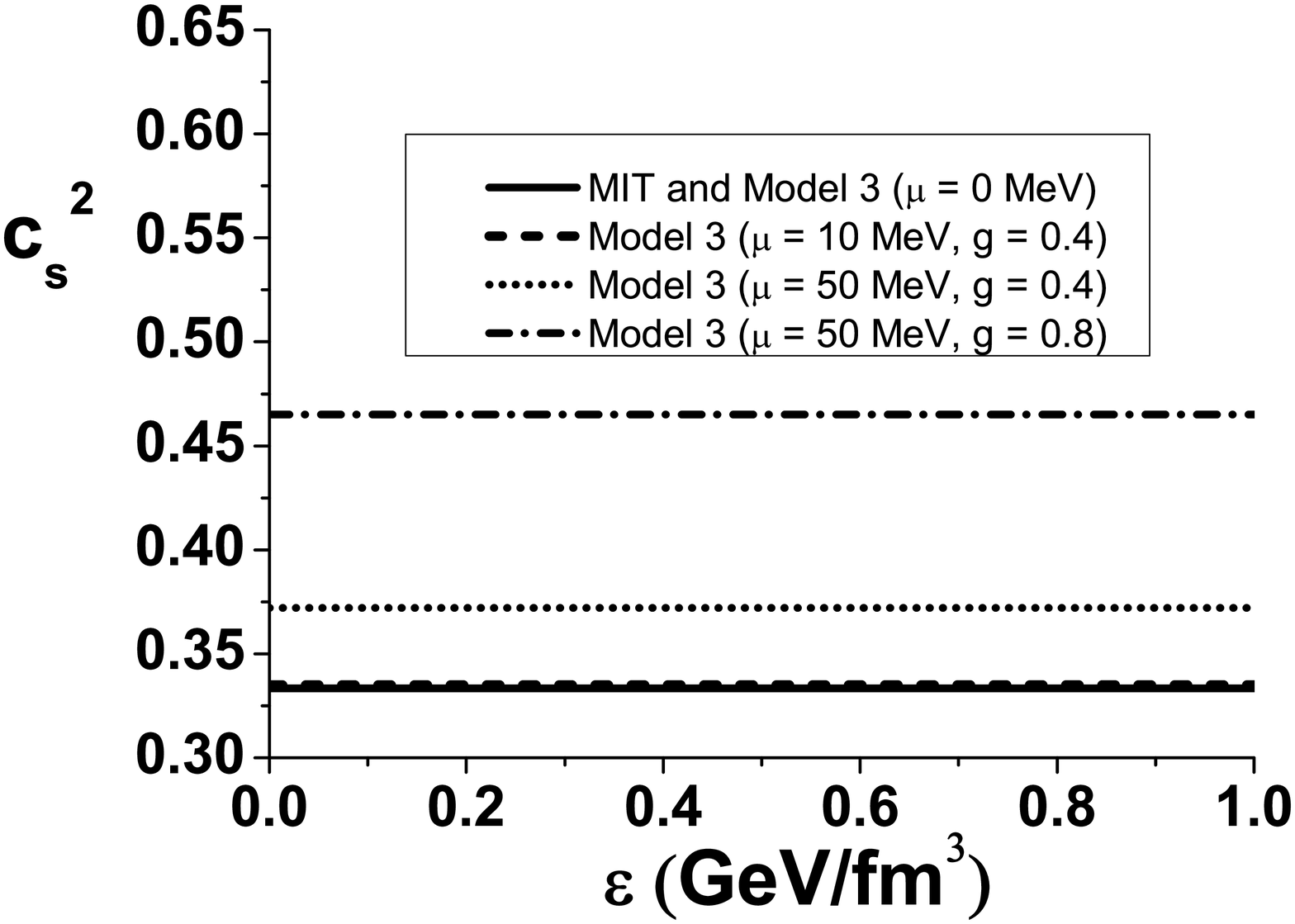}}
\subfigure[ ]{\label{fig:third}
\includegraphics[width=0.488\textwidth]{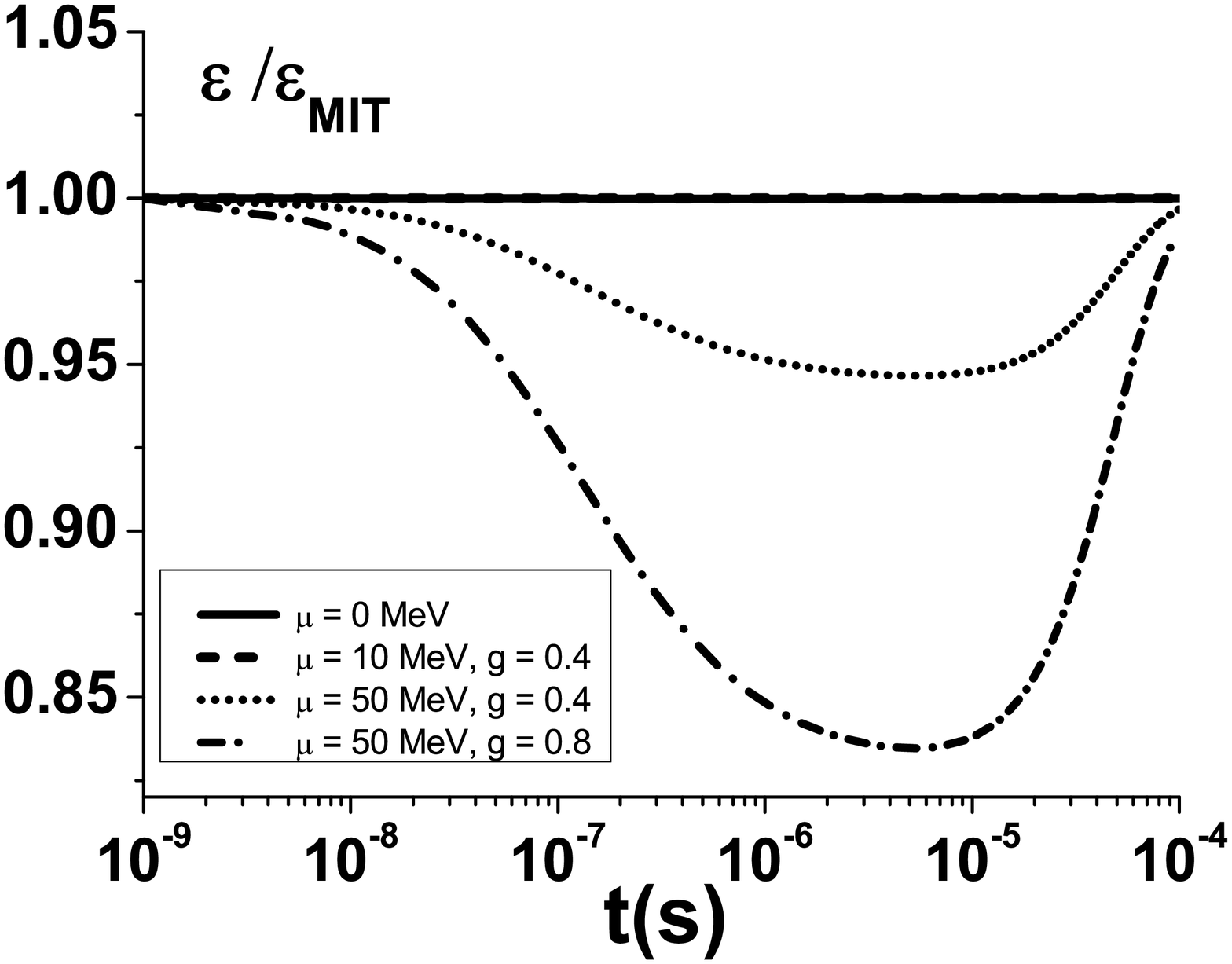}}
\subfigure[ ]{\label{fig:fourth}
\includegraphics[width=0.488\textwidth]{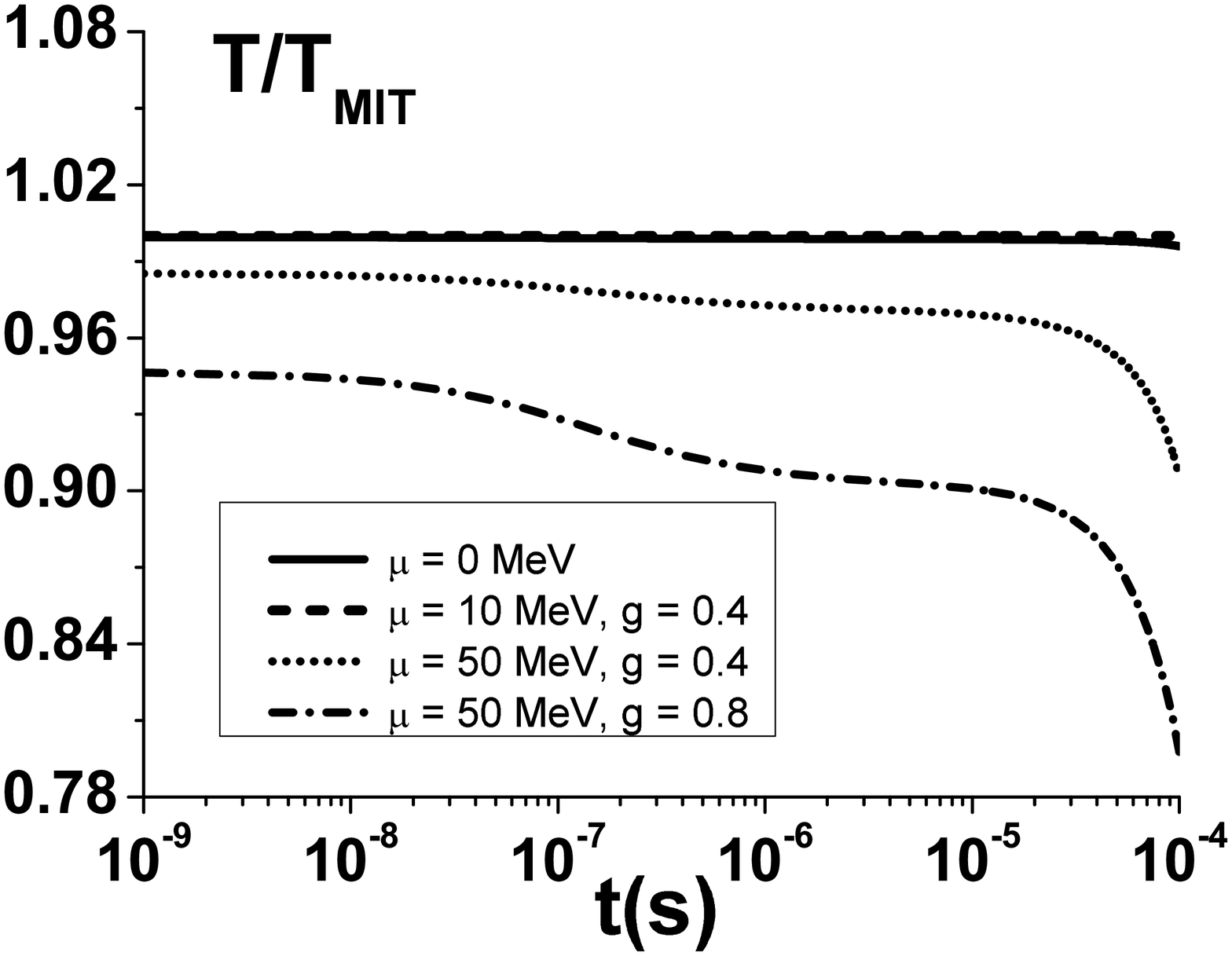}}
\end{center}
\vskip0.5cm
\caption{Comparison between model 3 $(m_{g}=10 \, MeV)$ and the MIT bag model. (a) Equation of state. (b) Speed of sound. (c) Time evolution of the energy density ratio.
(d) Time evolution of the temperature ratio.}
\label{fig3}
\end{figure}
In most of the cases, model 3  yields an equation of state which is stiffer  than the  MIT one. Looking at the expressions (\ref{phigh}) and (\ref{ehigh})
it is easy to see qualitatively  what are the effects of changing the gluon mass, the coupling constant and the chemical potential.
From these expressions we can conclude that model 3 will give always a harder equation of state, except when $\mu = 0$, in which case it reproduces
the bag model results. The numerical evaluation of (\ref{phigh}) and (\ref{ehigh}), presented in Fig. 3, shows that even taking a very small value of $m_g$,
a relatively large value for $g$ and a sizeable chemical potential we  obtain a speed of sound, which is  only 50 \% larger than the MIT value. Comparing
Figs. 2a and 3a, we notice that, the lattice EOS has always  less pressure whereas the mean field QCD EOS has always more pressure than the MIT bag model.  This
last feature is
necessary to reproduce the  available data on masses of compact stars \cite{demo}. Models 1 and 2 were tuned to fit the lattice data and the parameters of
model 3  were fixed at large $\mu$ and very small $T$ (so as to generate massive stars). The latter should also suffer some adjustments to reproduce also the
lattice results, taken at a quite different corner of the phase diagram with small $\mu$ and very large $T$. Compatibility between the two sets of data seems to
require that the speed of sound decreases with decreasing chemical potential. As it can be seen in Fig. 3a, this is already happening but the  reduction of
$c_s^2$ (which we can guess from the slopes in the figure) should be stronger. For the purposes of the present study we shall keep the parameters of model 3 in
the range determined in previous works.

The time evolution of the energy density and temperature of matter described by model 3 are shown in Fig. 3c and Fig. 3d respectively.
As it could have been anticipated from the previous figures, there is only a small difference between model 3 and the MIT bag model and this difference tends to
decrease with time in the case of the energy density and to remain approximately constant in the case of the temperature.

\subsection{Lattice QCD models}

The two lattice models (models 4 and 5) differ mainly because in the former there are only gluons whereas in the latter quarks are included.
We can hence identify the effect of the quarks in the equation of state and in the time evolution as well.
We first compare the results of  models $4$ and $5$, given by (\ref{lattpress}), (\ref{lattener}), (\ref{rattipre}) and (\ref{rattiener}), with the
EOS of a gas of   gluons  (\ref{eandpmitg}) and with the bag model EOS (\ref{eandpmit}). The time evolution of the  energy density is shown in  Fig. 4a.
As it can be seen there is only a small difference when we add quarks. In fact these models have nearly the same energy density during most of
the evolution and a sizeable difference appears only at very late times, already in the phase transition region.

In Fig. 4b we show the time evolution of the temperature calculated with the same models. It should
be noted that model 4 has its critical temperature at $T = 280$ MeV. In  model 5  the quark-hadron transition occurs at
$T = 200$ MeV.  In the figure we can see that for similar values of the energy density the models with more degrees of freedom have a smaller
temperature, as we would expect from simple thermodynamical considerations. In the phase transition region, the ratios change significantly and we may
expect that the use of the lattice models will yield important corrections to previous calculations made with the bag model.
\begin{figure}[ht!]
\begin{center}
\subfigure[ ]{\label{fig:first}
\includegraphics[width=0.488\textwidth]{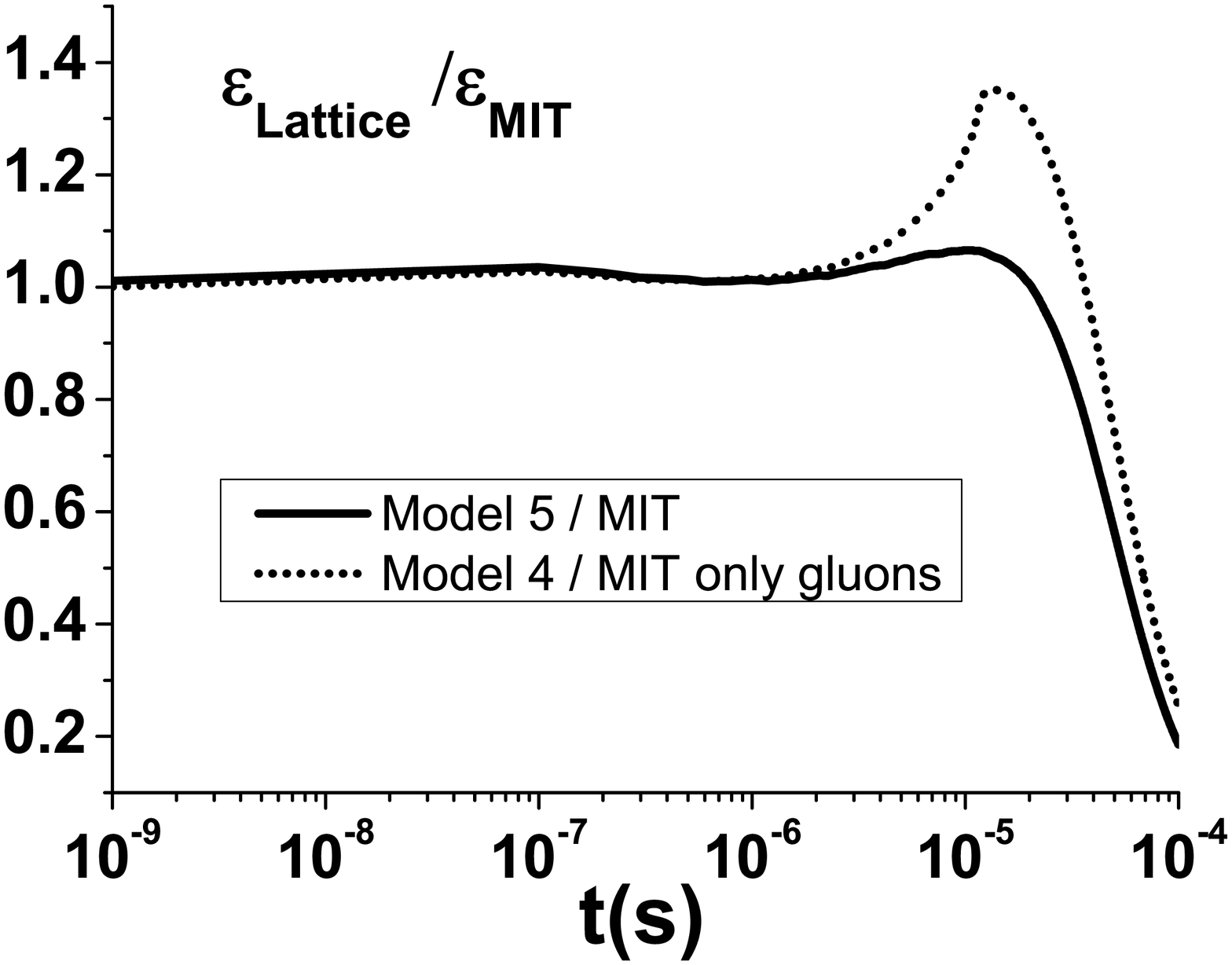}}
\subfigure[ ]{\label{fig:second}
\includegraphics[width=0.488\textwidth]{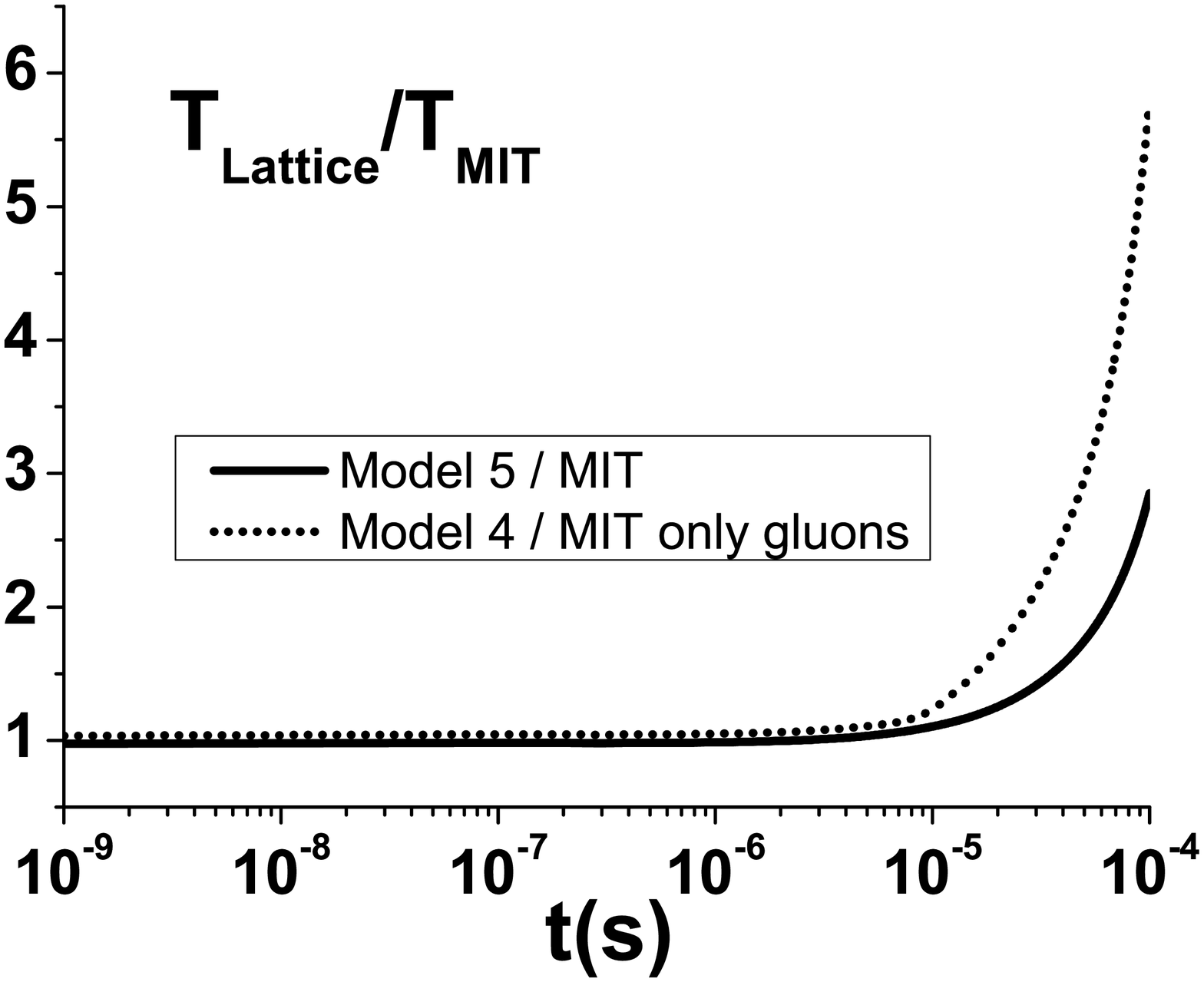}}
\end{center}
\caption{Comparison between model 4, model 5 and the MIT bag model. (a) Time evolution of the energy density.  (b) Time evolution of the temperature.}
\label{fig4}
\end{figure}

In Figs. 5a and 5b we show the time evolution of the energy density and temperature respectively, computed with model 5.
In the figures $\varepsilon(t)$ and $T(t)$ are scaled by the corresponding values of the energy density and temperature at zero chemical
potential. From the figures we can conclude that, in the range considered,  the
chemical potential does not affect the time evolution.
In Fig. 5c and Fig. 5d we compare the results of  model 3 and model 5, the two models with which we can obtain results at finite chemical
potential. The results are very similar to those shown in Fig. 4. As in the case of the bag model differences appear only at late times, but they can be
significant.
\begin{figure}[ht!]
\begin{center}
\subfigure[ ]{\label{fig:first}
\includegraphics[width=0.488\textwidth]{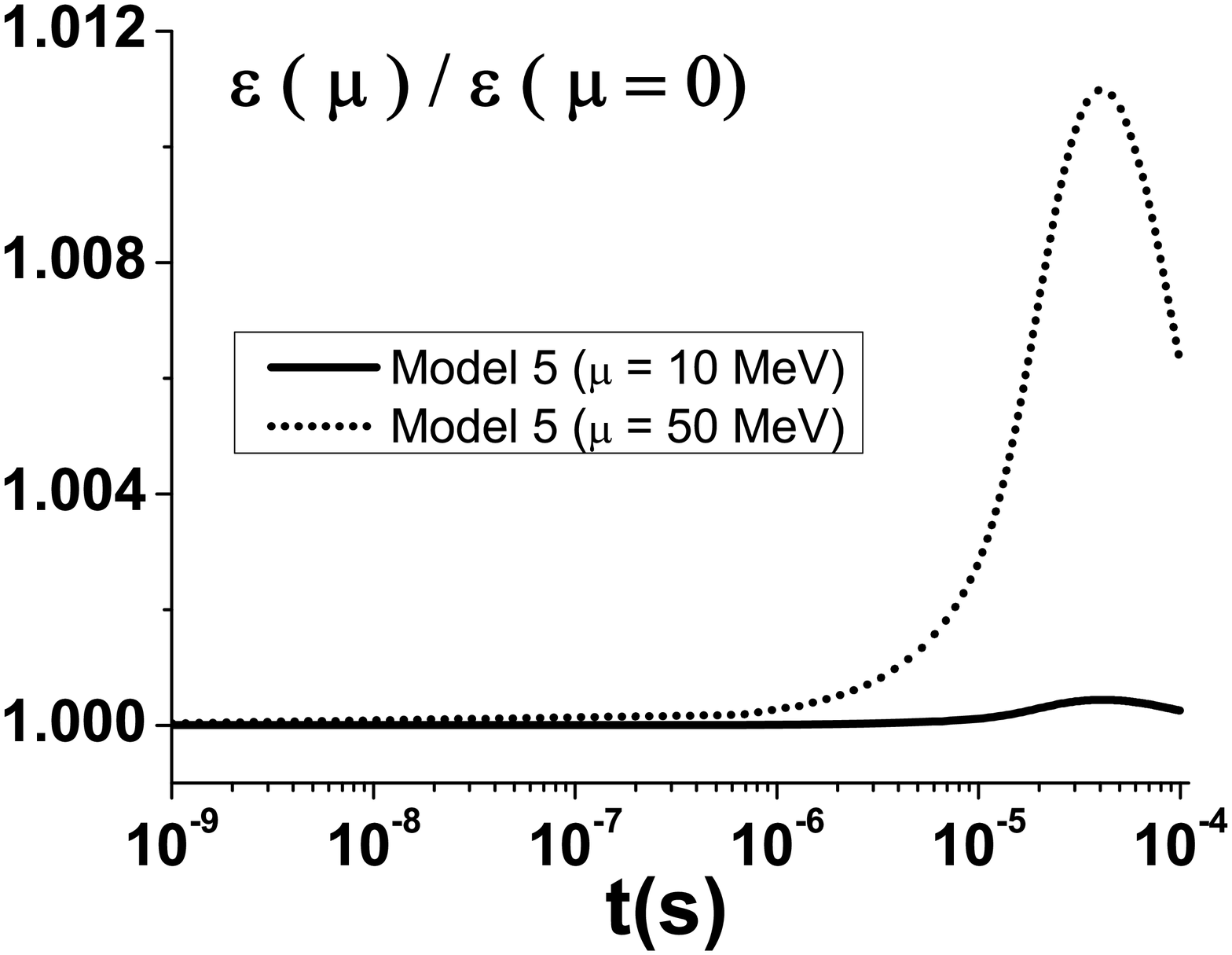}}
\subfigure[ ]{\label{fig:second}
\includegraphics[width=0.488\textwidth]{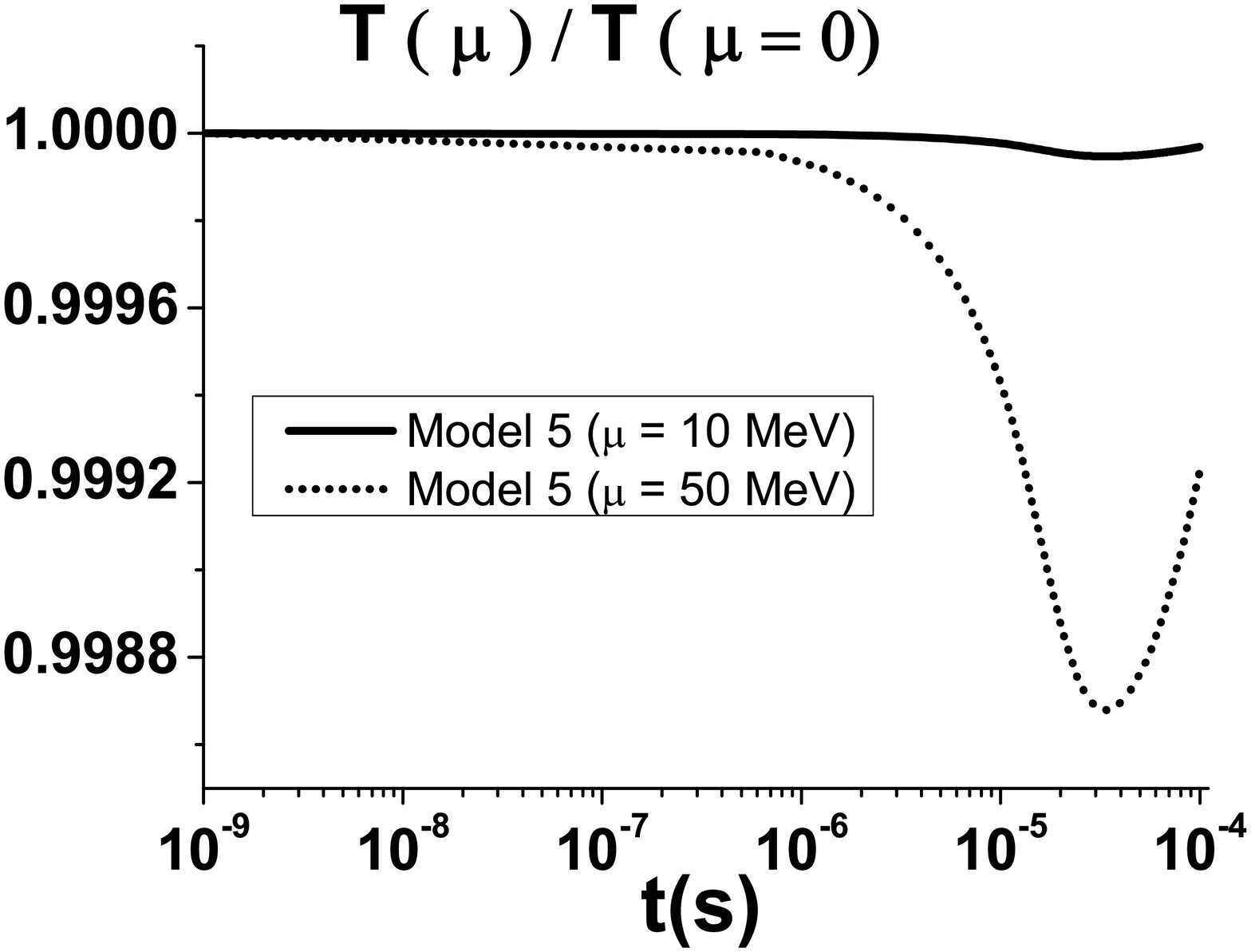}}
\subfigure[ ]{\label{fig:third}
\includegraphics[width=0.488\textwidth]{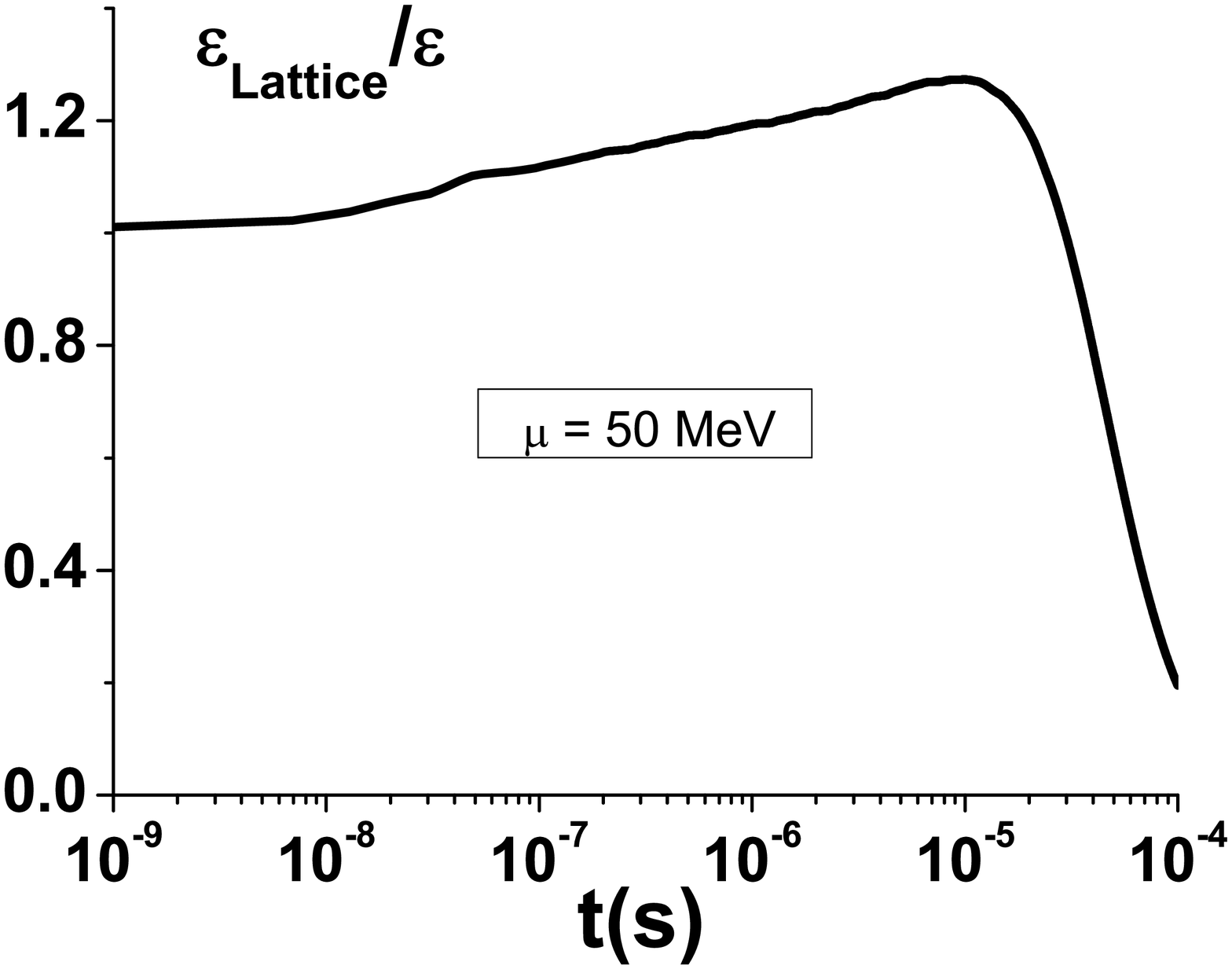}}
\subfigure[ ]{\label{fig:fourth}
\includegraphics[width=0.488\textwidth]{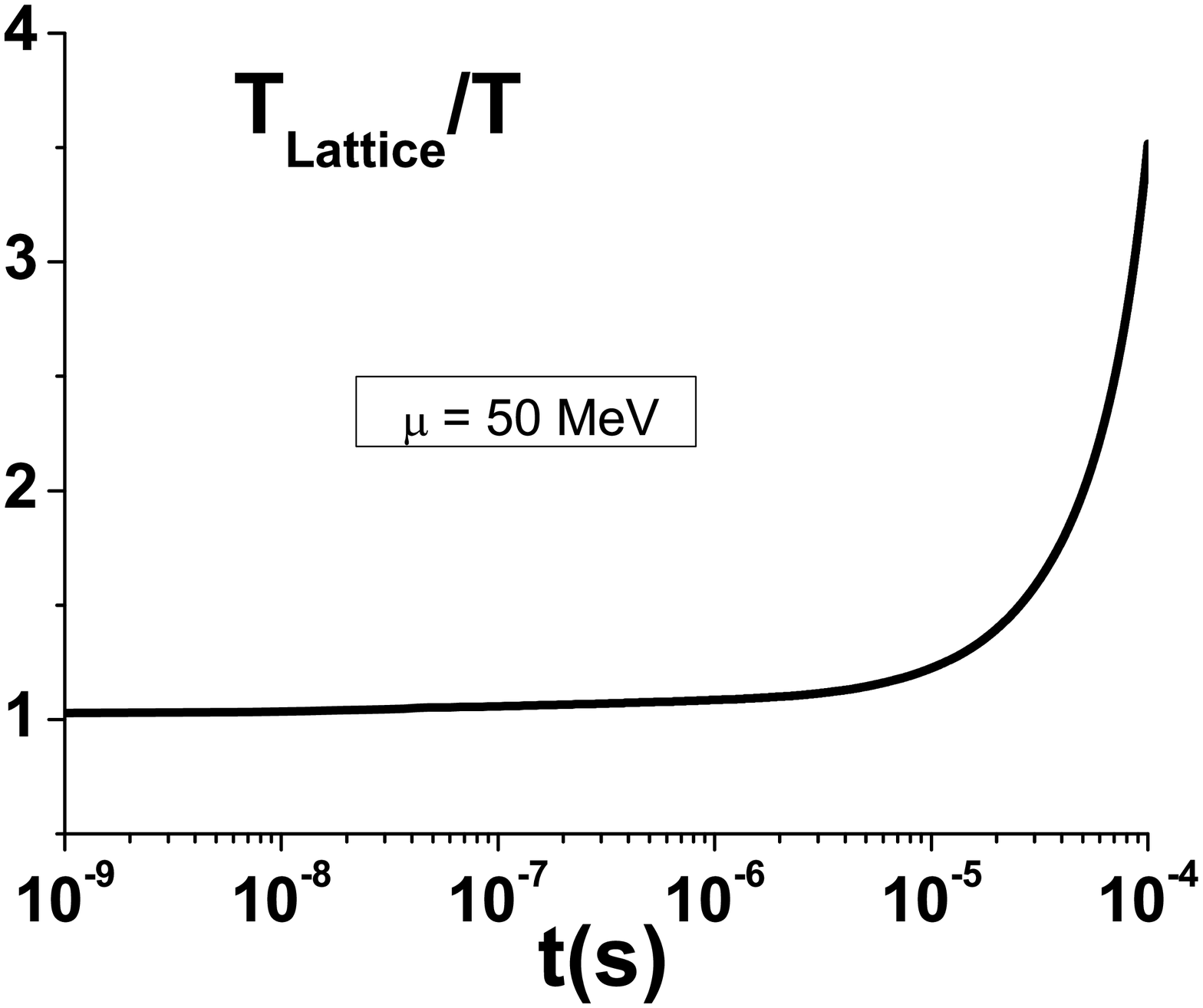}}
\end{center}
\caption{Comparison between model 5 and model 3 (with $m_{g} = 10 $ MeV and $g = 0.8$).
(a) Effect of a finite chemical potential on the time evolution of the energy density. (b)  Effect of a finite
chemical potential  on the  time evolution of the temperature.  (c) Ratio of the energy densities in model 5 and model 3  as a function of time.
(d) Ratio of the temperatures in model 5 and model 3 as a function of time.}
\label{fig5}
\end{figure}

\subsection{The electroweak contribution}

\begin{figure}[ht!]
\begin{center}
\subfigure[ ]{\label{fig:first}
\includegraphics[width=0.488\textwidth]{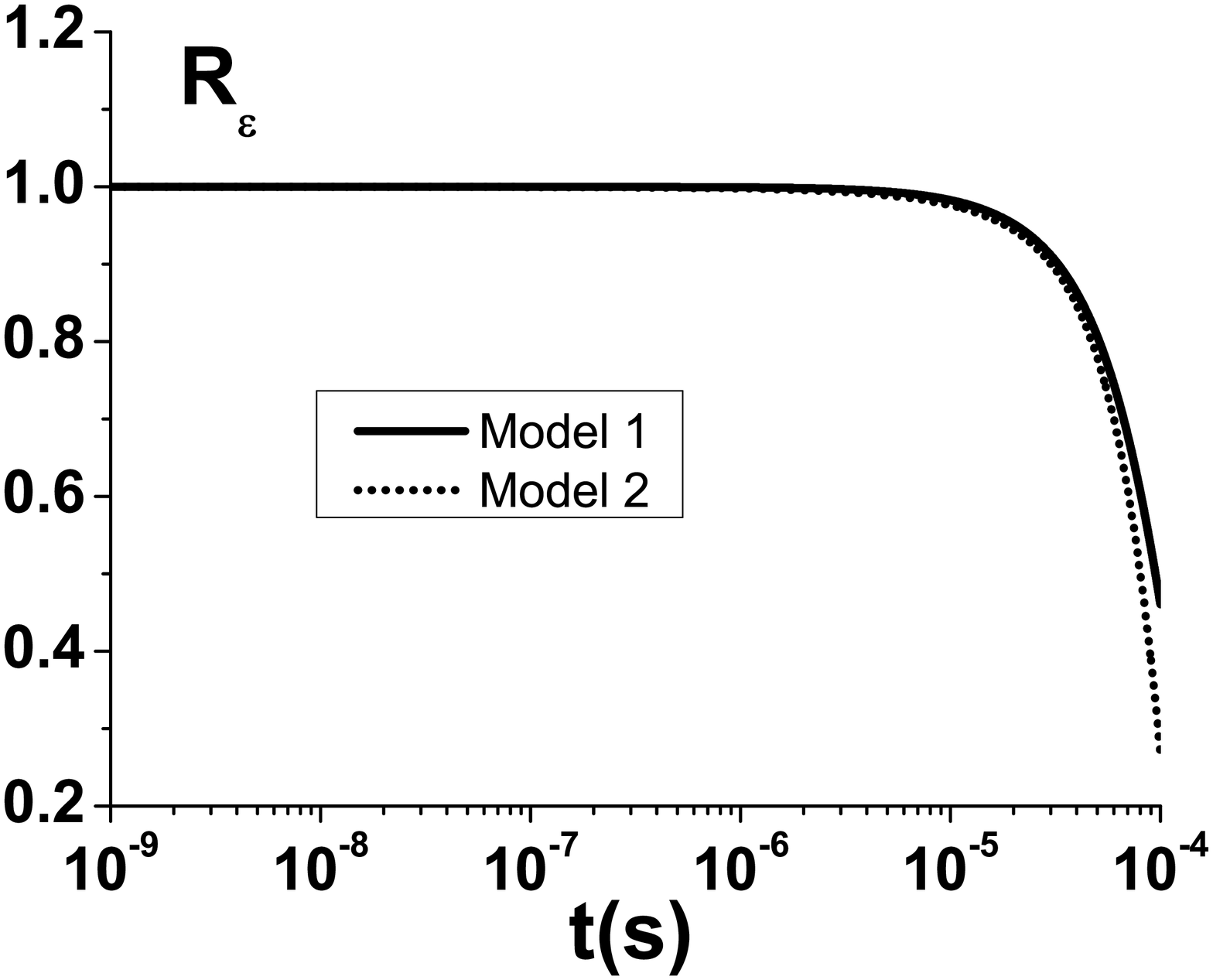}}
\subfigure[ ]{\label{fig:second}
\includegraphics[width=0.488\textwidth]{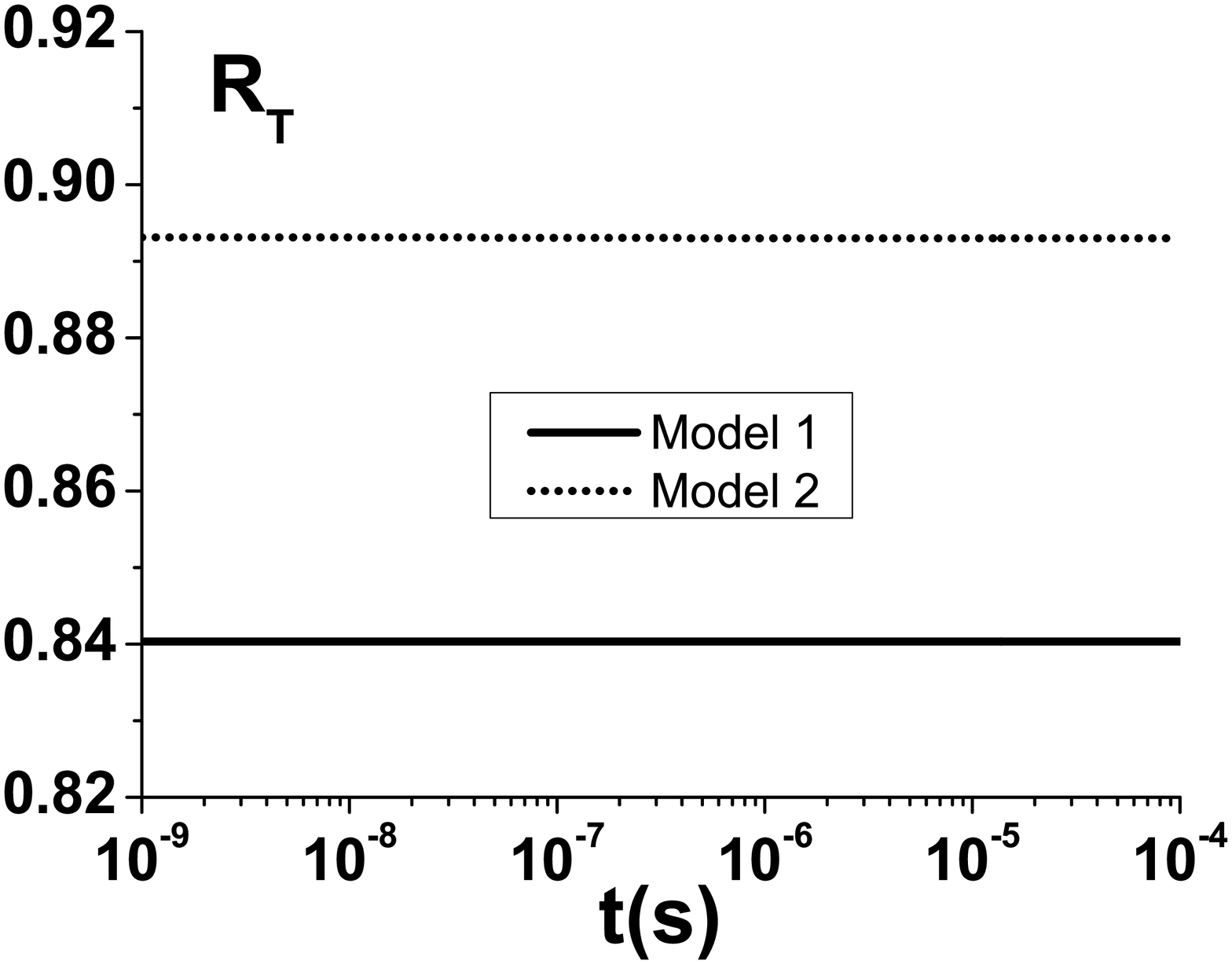}}
\subfigure[ ]{\label{fig:third}
\includegraphics[width=0.488\textwidth]{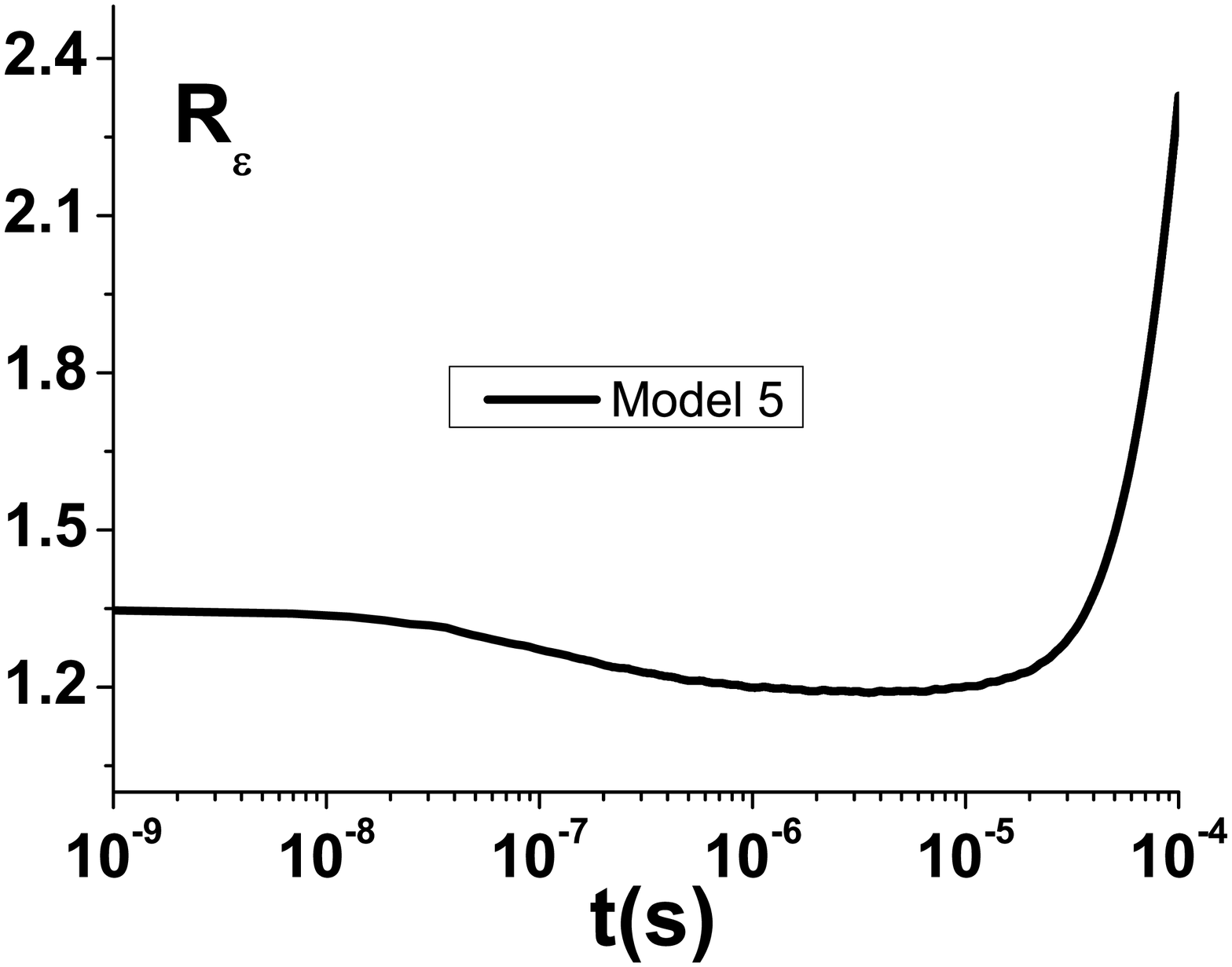}}
\subfigure[ ]{\label{fig:fourth}
\includegraphics[width=0.488\textwidth]{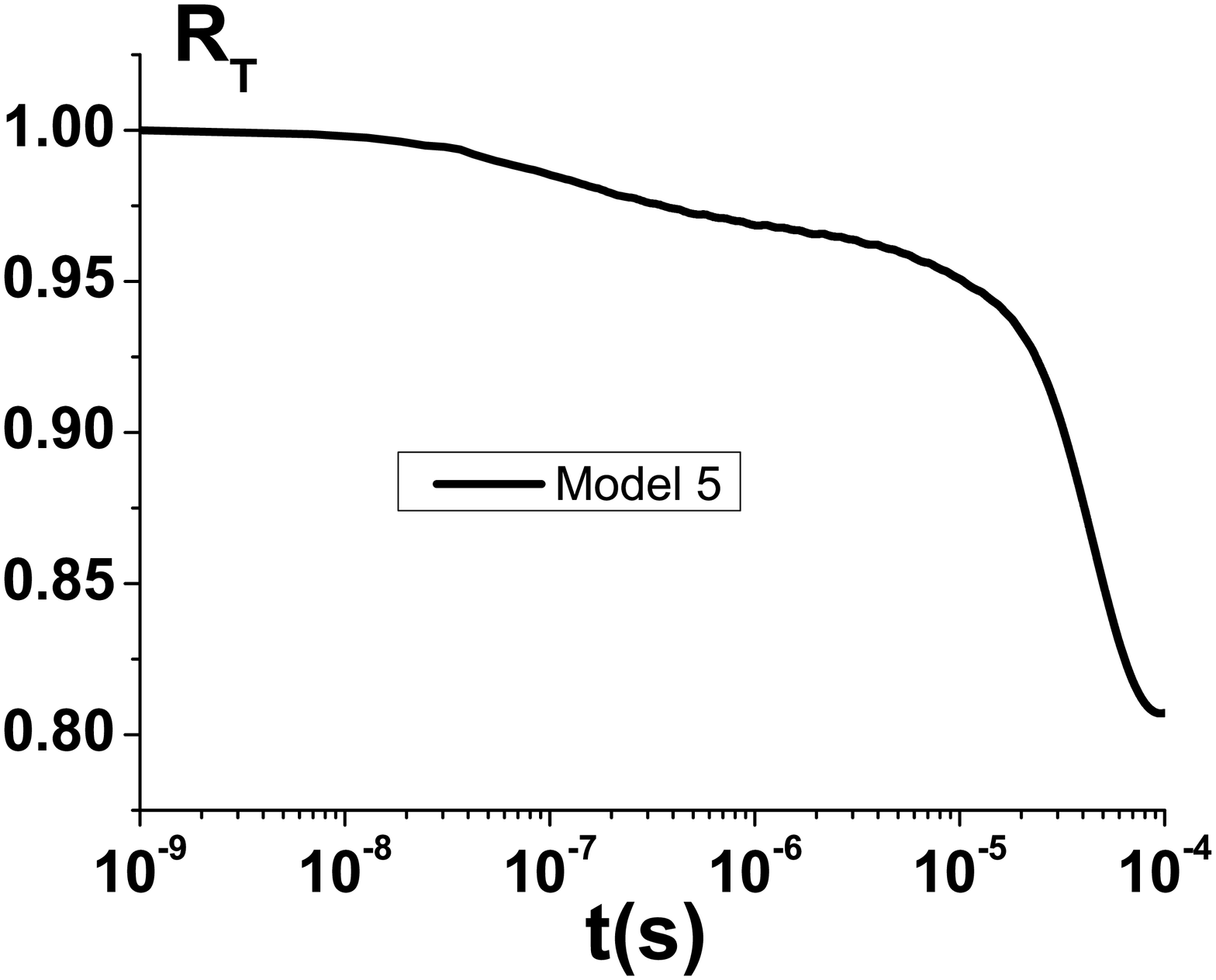}}
\end{center}
\caption{Effect of the electroweak contribution on the time evolution of models 1,  2 and 5.
Ratio of the energy density (and temperature) of the full models and the models without the EW contribution as a function of time.
(a) $R_{\varepsilon}$ for model 1 and model 2.  (b) $R_T$ for model 1 and model 2.
(c)  $R_{\varepsilon}$ for model 5. (d) $R_T$  for model 5.}
\label{fig6}
\end{figure}

As in previous works, we describe the plasma in the early Universe as a quark-gluon plasma plus electroweak matter in thermal
equilibrium. The presence of the electroweak matter component (EW) is one of the remarkable  differences between the primordial QGP
and the QGP produced in heavy ion collisions and it is very important to determine its effect on the thermodynamical properties
and on the expansion of the early Universe. The electroweak matter has energy density and pressure given by  \cite{japas,guardo,florko}:
\begin{equation}
\varepsilon_{ew} = g_{ew} \frac{\pi^2}{30} \,T^{4}
\hspace{1.5cm} \textrm{and} \hspace{1.5cm}
p_{ew} = g_{ew}  \frac{\pi^2}{90} \,T^4
\label{ew}
\end{equation}
where $g_{ew} = 14.45$.
The total energy and pressure of the system are then given by:
\begin{equation}
\varepsilon = \varepsilon_{i} +  \varepsilon_{ew}
\hspace{1.5cm} \textrm{and} \hspace{1.5cm}
p = p_i + p_{ew}
\label{epfin}
\end{equation}
where $\varepsilon_{i}$ and $p_i$ are the energy and pressure of the QGP given by model $i$, respectively. From (\ref{eandpmit}) and (\ref{ew})
we see that for any given temperature the QGP contains much more (a factor of $\simeq 2.5$) energy and pressure than the electroweak matter.
In Fig. \ref{fig6} we illustrate the effects of the EW component on the evolution of the QGP described by MIT-based models  (model 1 and model 2)
and by the lattice (model 5).  We solve the Friedmann equation (\ref{eq:time-evol-E/V}) using $\varepsilon = \varepsilon_{i} +  \varepsilon_{ew}$
and $p = p_i + p_{ew}$ for $i=1,2$ and $5$. These are called the ``full models''. Then we solve it again using  $\varepsilon = \varepsilon_{i}$ and  $p = p_i$,
compute the ratios of the two solutions and plot them in  Fig. \ref{fig6}. We define $R_{\varepsilon} = \varepsilon(t)/\varepsilon_{i}(t)$ and
$R_{T} = T(t)/T_{i}(t)$. The initial condition (\ref{init}) implies that all plasmas start with the
same energy density and hence the energy ratio starts being one, whereas the temperature ratio is smaller than one and depends on the
details of the equation of state.  For systems with a larger number of degrees of freedom at the same energy density, the temperature
should be smaller, as we can see in  Figs. \ref{fig6}b and    \ref{fig6}d. From these figures we can conclude that
the EW does not affect the time evolution of the QGP except at very late stages, when the system approaches the quark-hadron phase
transition. At this point we see a that the MIT variants and  the lattice model 5 have a qualitatively different  behavior.

\subsection{Scale factor}

\begin{figure}[ht!]
\begin{center}
\subfigure[ ]{\label{fig:first}
\includegraphics[width=0.488\textwidth]{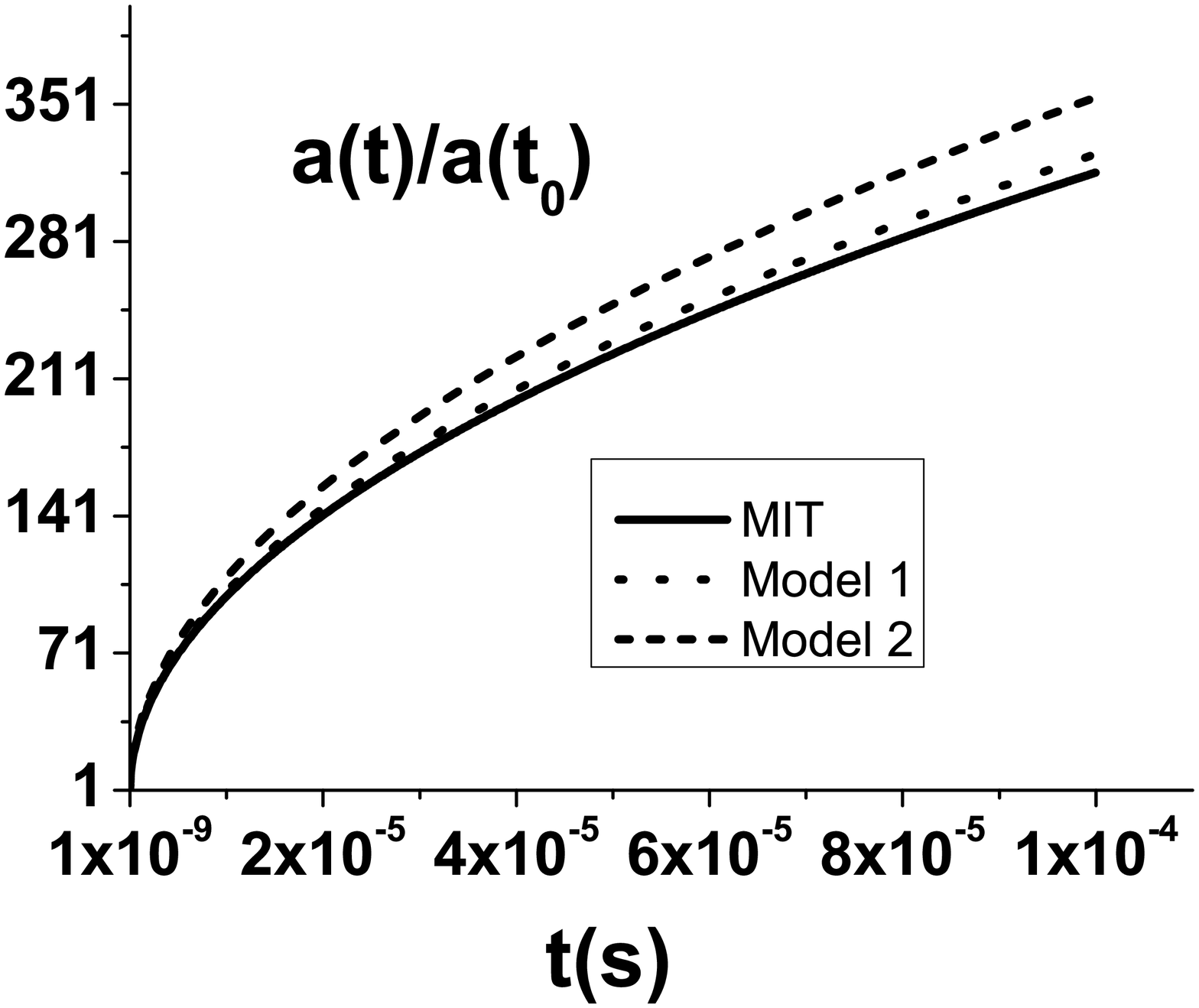}}
\subfigure[ ]{\label{fig:second}
\includegraphics[width=0.488\textwidth]{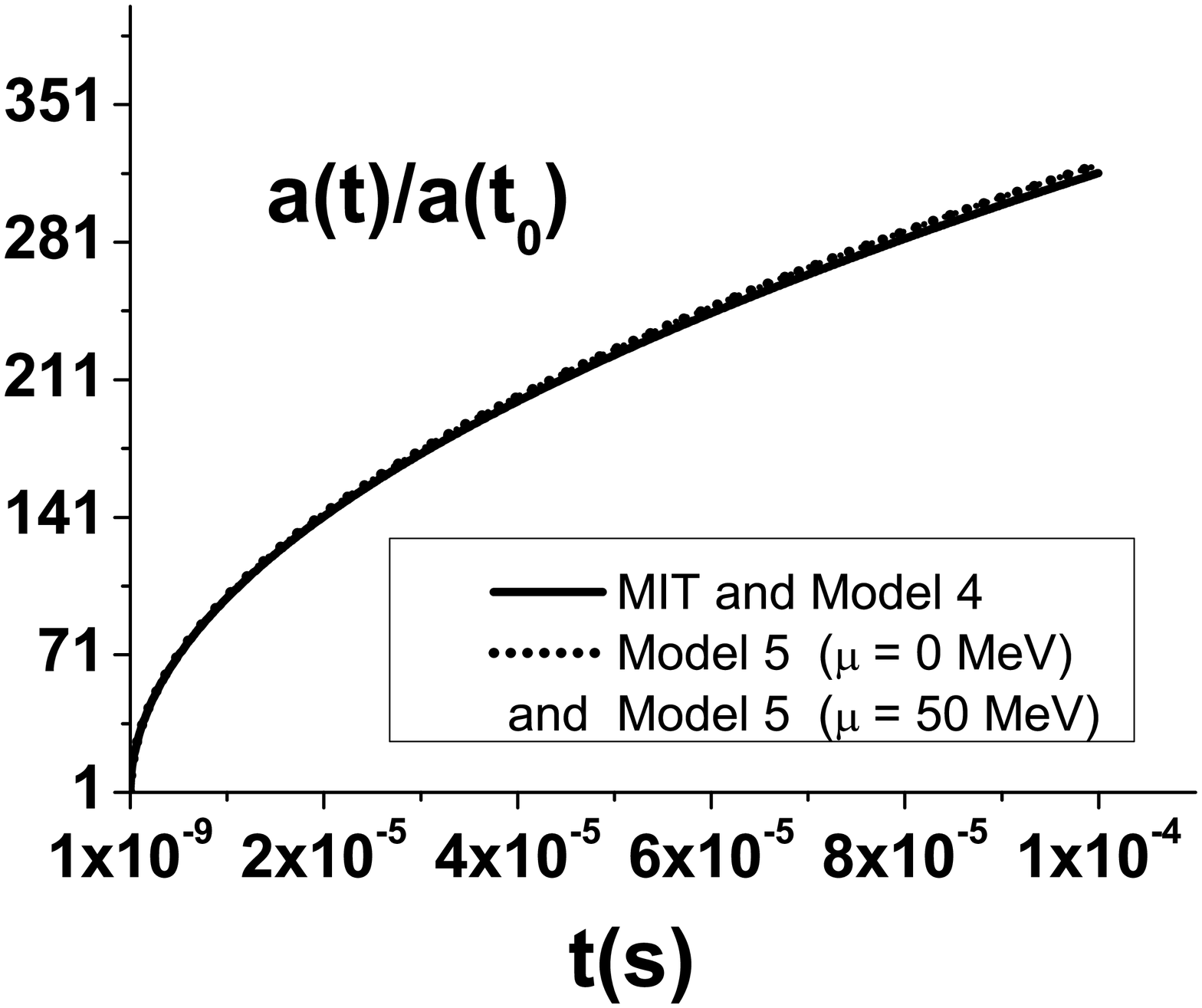}}
\subfigure[ ]{\label{fig:first}
\includegraphics[width=0.488\textwidth]{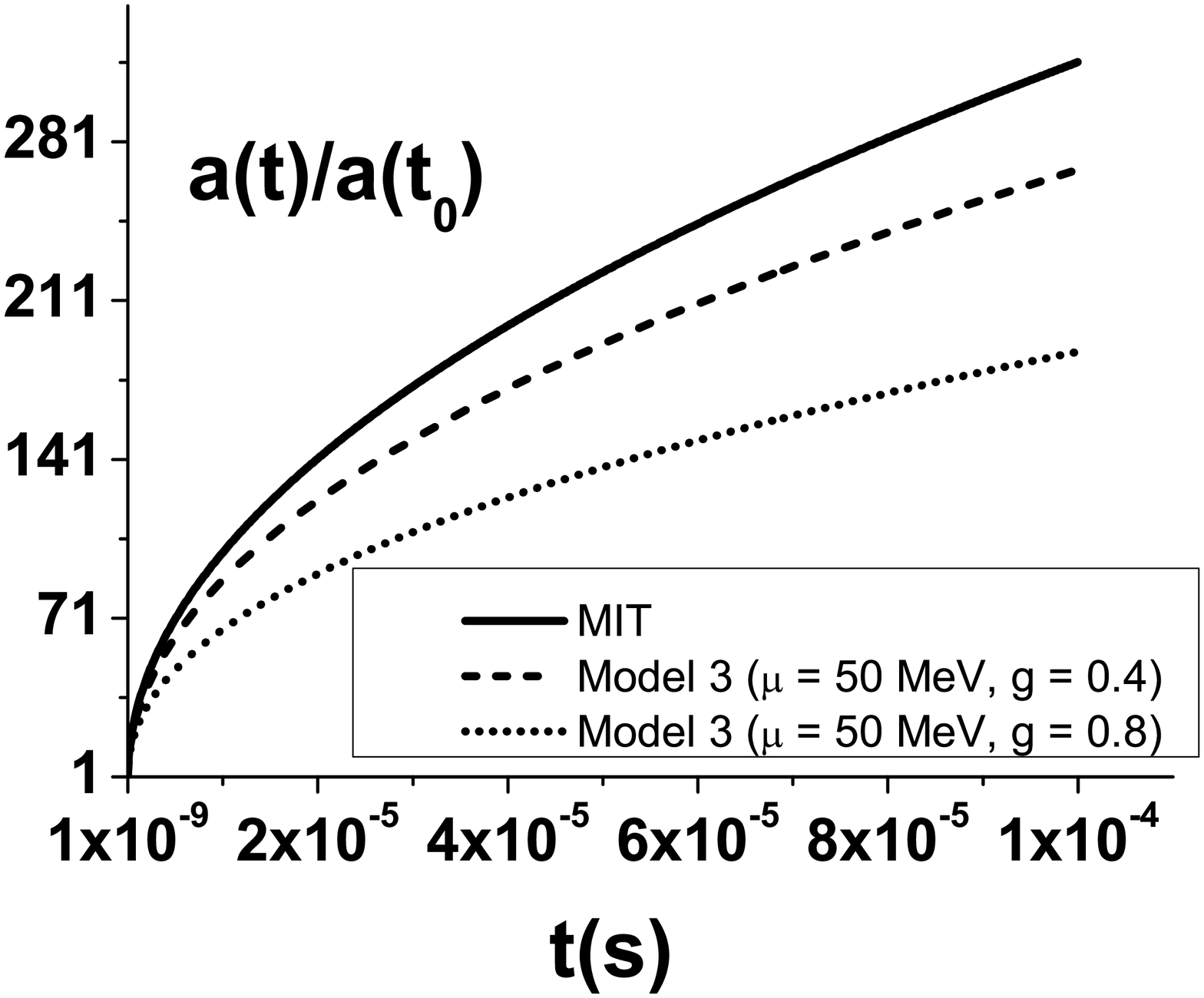}}
\subfigure[ ]{\label{fig:second}
\includegraphics[width=0.488\textwidth]{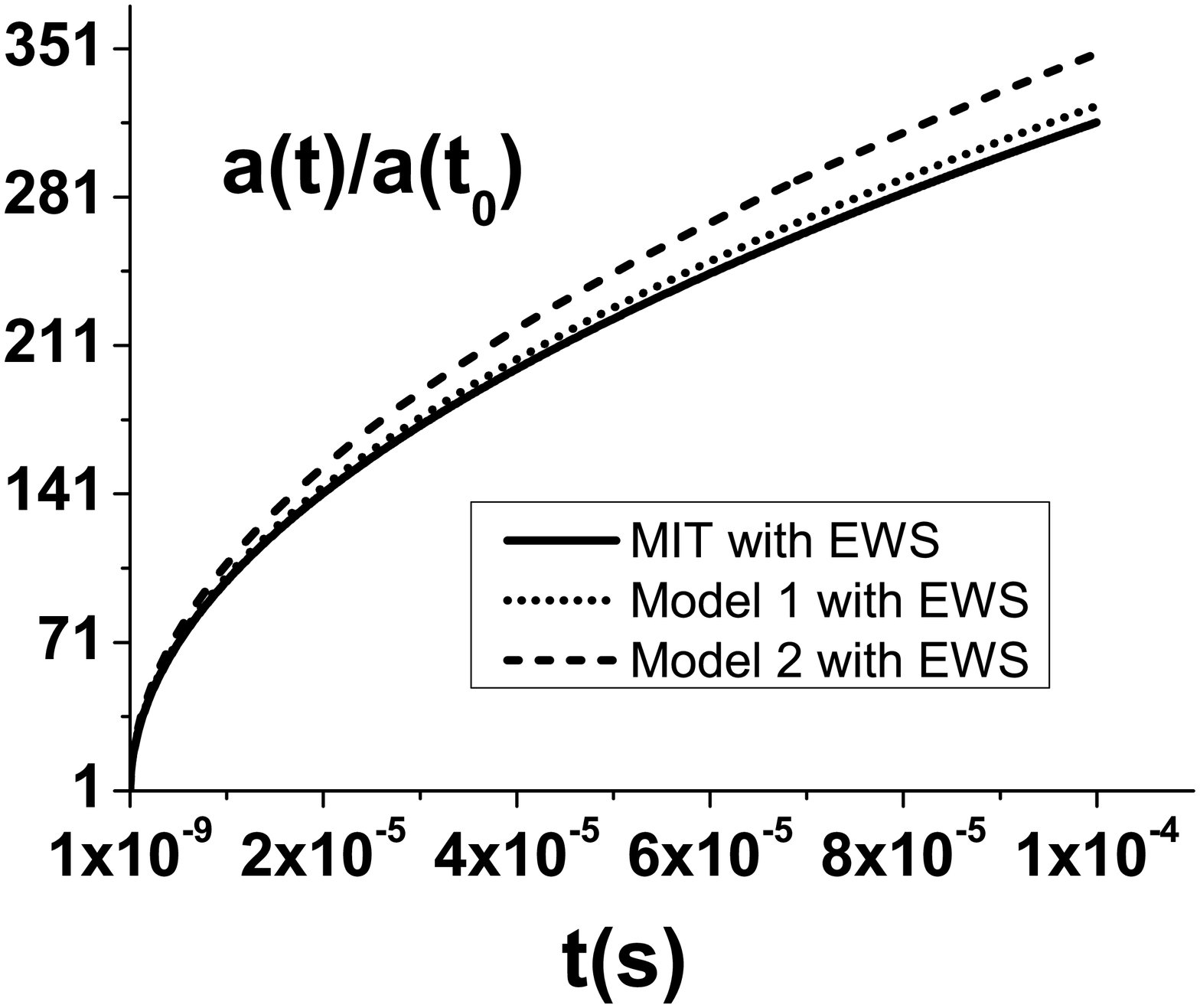}}
\end{center}
\caption{Scale factor as a function of time given by Eq. (\ref{ratioscale}). (a) MIT based equations of state. (b) Lattice based equations of state.
(c) Mean field QCD equation of state (Model 3 with $m_{g}=10 \, MeV)$.  (d) Effect of the EW component.}
\label{fig7}
\end{figure}

As in  \cite{japas,guardo,florko,cosmo2}  we can write the Friedmann equation in the following form:
\begin{equation}
{\frac{\dot{a}(t)}{a(t)}}=
-\frac{\dot{\varepsilon}(t)}{3 \Big[ \varepsilon(t) + p(t) \Big]} = \sqrt{\frac{8 \pi G}{3} \varepsilon(t)}
\label{fried}
\end{equation}
where $a(t)$ is the scale factor. Let us consider the time interval from  the birth of the Universe, $t=0$, to the ``initial time''
of our interest: $t_0$.  The corresponding energy densities are $\varepsilon(t=0) \equiv \tilde{\varepsilon}$ and
$\varepsilon(t=t_{0}) \equiv {\varepsilon}_{0}$ respectively. The energy density at the beginning of the Universe is huge and hence
$\tilde{\varepsilon} >> {\varepsilon}_{0}$. Following \cite{florko} we shall assume that the sound speed is constant in time and
therefore $p(t)={c_{s}}^{2}\,\varepsilon(t)$. The second equality in (\ref{fried}) can be easily integrated and then, solving it for
$t_0$ and taking $(1/\sqrt{\tilde{\varepsilon}}) \, \rightarrow 0$, we find:
\begin{equation}
t_{0}={\frac{1}{\sqrt{6 \pi G(1+{c_{s}}^{2})^{2}\,\varepsilon_{0}}}}
\label{tzero}
\end{equation}
We now consider another time interval from the initial time $t_0$, when the energy density is  $\varepsilon_{0}$,  to a generic time $t$
where  the energy density is  $\varepsilon $. Integrating again the second equality of (\ref{fried}) in this time interval, using (\ref{tzero})
and solving for  $\varepsilon(t)$ we find:
\begin{equation}
\varepsilon(t)={\frac{1}{6 \pi G(1+{c_{s}}^{2})^{2}\,t^{2}}}
\label{epst}
\end{equation}
Knowing the time dependence of the energy density it is a simple exercise to go back to  (\ref{fried}), integrate it from
$t_0$ to time $t$ and find:
\begin{equation}
{\frac{a(t)}{a(t_{0})}}=\Bigg({\frac{t}{t_{0}}}\Bigg)^{{\frac{2}{3(1+{c_{s}}^{2})}}}
\label{ratioscale}
\end{equation}
In \cite{florko} this same expression  was obtained for the particular case where $c_s^2 = 1/3$ and thus $a/a(t_{0})=(t/t_{0})^{1/2}$.
For the sake of completeness, we show in Fig. \ref{fig7} the ratio $a/a(t_{0})$ as a function of time. Taking the
predictions of the MIT bag model as a reference we compare the evolution of the scale factor computed with the  MIT variants
(Fig. \ref{fig7}a), with the lattice models 4 and 5 (Fig. \ref{fig7}b) and with the mean field model 3
(Fig. \ref{fig7}c). In Fig. \ref{fig7}d
we show the effect of the inclusion of the EW component on the evolution of the scale factor. As it is clear from Eq. (\ref{ratioscale}),
the only differenc in these curves comes from the speed of sound, which changes from model to model but is constant in time. The limiting cases
are easy to identify: when $c_s^2 = 0$ (dust) the expansion is faster and when the speed of sound reaches its upper limit, $c_s^2 = 1$, the
expansion is slower.  The behavior observed in Figs. \ref{fig7}a and  \ref{fig7}c is a direct consequence of Figs. 2a and 3a
respectively. In the cases considered here, larger pressures led to larger $c_s^2$ and thus to slower expansion rates.

\section{Conclusions}

As it was mentioned in the introduction, in view of the future experimental facilities for  measurements relevant to  cosmology,
it is very interesting and very timely to identify observable effects of the QGP phase in the primordial Universe.  Most of these effects have so far
been associated with quark-hadron phase transition. However we think that it is also worth studying the time evolution of the QGP phase and check whether
changes in the QGP equation of state induce changes in the dilution and cooling of the early Universe.  These changes might, in turn, change the emission of
gravitational waves or the  generation of baryon number fluctuations. Taking the equation of state of the MIT bag model as
a baseline, we have considered other EOS with different dynamical ingredients and estimated their effects on the solutions of Friedmann equations. The main
conclusion is that there are no dramatic changes in the whole time interval considered. The time evolution of the energy density is only weakly
sensitive to changes in the chemical potential, to changes in the degrees of freedom (addition of quarks) and to changes in the dynamical information encoded
in the effective gluon mass. The temperature evolution is somewhat more sensitive to these changes, specially to the inclusion of quarks in a pure gauge theory.
Although our results are already very suggestive, our preliminary conclusions need to be confirmed by a more complete calculation,
with the inclusion of the phase transition and the evolution of the hadronic phase. Finally, we emphasize that there are still some other ingredients of the
QGP phase to be considered, such as, for example the number of leptons and a strong magnetic field.  Calculations along this line are in progress.

\begin{acknowledgments}
This work was partially financed by the Brazilian funding agencies CAPES, CNPq and FAPESP. We thank C. Ratti, L.R.W. Abramo and G. Lugones for instructive discussions.
\end{acknowledgments}

\end{document}